\documentclass[preprintnumbers,showpacs,amsmath,amssymb]{revtex4}
\usepackage{epsfig}
\usepackage{graphicx}  
\usepackage{color}     
\usepackage{amscd}
\usepackage{color}
\usepackage{lscape}
\usepackage{subfigure}
\date{\today}
 \normalsize

\newcommand{\bmat}{\left(\begin{array}}
\newcommand{\emat}{\end{array}\right)}
\newcommand{\be}{\begin{equation}}
\newcommand{\ee}{\end{equation}}
\newcommand{\bea}{\begin{eqnarray}}
\newcommand{\eea}{\end{eqnarray}}

\def\gtwid{\mathrel{\raise.3ex\hbox{$>$\kern-.75em\lower1ex\hbox{$\sim$}}}}
\def\ltwid{\mathrel{\raise.3ex\hbox{$<$\kern-.75em\lower1ex\hbox{$\sim$}}}}
\def\gev{{\rm \, Ge\kern-0.125em V}}
\def\tev{{\rm \, Te\kern-0.125em V}}

\def    \be            {\begin{equation}}
\def    \ee            {\end{equation}}
\def    \bea           {\begin{eqnarray}}
\def    \eea           {\end{eqnarray}}

\def\a{\alpha}

\def\d{\delta}

\def\om{\omega}

\def\sig{\sigma}

\def\th{\theta}

\def\d{\delta}

\def\la{\lambda}
\def\nn{\nonumber}
\begin{document}
\title{Two driven coupled qubits in a time varying magnetic field: exact and approximate solutions}
\author{E. I. Lashin}
\affiliation{Department of Physics, Ain Shams University, Cairo 11566, Egypt and \\
Centre for Theoretical Physics, Zewail City of Science and Technology,
Sheikh Zayed, 6 October City, 12588, Giza, Egypt.}
\author{Gehad Sadiek}
\affiliation{Department of Physics, King Saud University, Riyadh 11451, Saudi Arabia and\\
Department of Physics, Ain Shams University, Cairo 11566, Egypt}
\author{M. Sebaweh Abdullah}
\affiliation{Department of Mathematics, King Saud University, Riyadh 11451, Saudi Arabia}
\author{Elham Aldufeery}
\affiliation{Department of Physics, King Saud University, Riyadh 11451, Saudi Arabia and\\
Department of physics, Majmaah University, Zulfi 11932, Saudi Arabia}
\begin{abstract}
We study the dynamics of a two-qubit system coupled through time dependent anisotropic $XYZ$ Heisenberg interaction in presence of a time varying non-uniform external magnetic field. Exact results are presented for the time evolution of the system under certain integrability conditions. Furthermore, the corresponding entanglement of the system is studied for different values of the involved parameters. We found that the time evolution and different properties of entanglement such as amplitude, frequency and profile can be finely tuned by the interplay of the characteristics of the time-dependent magnetic field and exchange coupling. Also we show how the discrete symmetries of the system Hamiltonian, which splits its Hilbert space into two distinct subspaces, can be utilized to deduce the dynamics in one of its two distinct subspaces from the other one. Moreover, approximate results for the time evolution are provided utilizing the rotating wave as well as the perturbation approximations for the special case of either static magnetic field or exchange coupling. We compare the range of validity of the two approximation methods and their effectiveness in treating the considered system and determine their critical parameters.\\
\end{abstract}
\maketitle
\section{Introduction}
Quantum entanglement is considered as the corner stone of the quantum theory and one of its historic puzzles. Nowadays it is considered as a well established concept and experimentally verified phenomenon in modern physics \cite{PeresBook}. Quantum entanglement is a nonlocal correlation between two (or more) quantum systems such that the description of their states has to be done with reference to each other even if they are spatially well separated.
Particular fields where entanglement plays a crucial role are quantum teleportation, quantum cryptography and quantum computing \cite{Nielsen, Bouwmeester, gruska, macchiavelleo}. Entanglement is considered as the physical basis for manipulating linear superposition of quantum states to implement the different proposed quantum computing algorithms \cite{Shor, Grover}. Different physical systems have been proposed as reliable candidates for the underlying technology of quantum computing and quantum information processing \cite{Barenco, ibm-stanford, NMR1, NMR3, TrappedIones, CvityQED, JosephsonJunction}. The main task in each one of these systems is to define certain quantum degree of freedom to serve as a qubit, such as the charge, orbital or spin angular momentum. The next step is to find a controllable mechanism for forming an entanglement between a two-qubit system in such a way to produce a fundamental quantum computing gate. In addition, we need to be able to coherently manipulate such as entangled state to provide an efficient computational process.
Such coherent manipulation of entangled state has been observed in systems such as isolated trapped ions \cite{trapped-ions} and superconducting junctions \cite{supercond-junc}. The coherent control of a two-electron spin state in a coupled quantum dot was achieved experimentally, where the coupling mechanism is the Heisenberg exchange interaction between the electron spins \cite{experim-Johnson, experim-Koppens, experim-Petta}.

Solid state systems have been in focus of interest as they can be utilized to build up integrated networks that can perform quantum computing algorithms at large scale.
Particularly, the semiconductor quantum dot is considered as one of the most promising candidates for playing the role of a qubit \cite{spin-qubit, spin-qgate, spin-orbit}, where the spin {\bf S} of the valence electron on a single quantum dot is used as a two-state quantum system, which gives rise to a well-defined qubit.
As a result of this especial interest, so much efforts has been devoted to investigating the interacting Heisenberg spin chain as it represents a very reliable model for constructing quantum computing schemes in different solid state systems, as well as for being a very rich model for studying the novel physics of localized spin systems \cite{spin-qgate, spin-orbit, nuclear-spins, optical-lattices}.

Recently there has been much of literature focusing on entanglement and decoherence in Heisenberg spin systems in presence and absence of an external magnetic field \cite{thermal-entanglement, spin-squeezing, phase-tansition, chain, spin-chain, disturbance, bendular, mean-field, information, dynamics, Sadiek2010-PRA82, Sadiek2011-PRA83,Sadiek2012-PRA84}. Particularly, the thermal entanglement of an isotropic two-qubit Heisenberg XY model under the effect of a non-uniform static magnetic field was studied, with an emphasis on the critical role of non-uniformity of the field on the entanglement \cite{multi}. The entanglement of anisotropic two-qubit Heisenberg XYZ model in presence of a non-uniform static magnetic field was investigated. It was demonstrated that the combined effect of the anisotropic coupling and the non-uniformity of the field has a significant impact on manipulating the details of the transition between the entangled and the disentangled states of the system \cite{anisotropic}. Furthermore the anisotropic two-qubit Heisenberg XYZ model was studied in presence of time-dependent magnetic fields, however, the magnetic field was assumed to be uniform \cite{traveling}. In previous works we have studied the dynamics of entanglement and thermal entanglement of a two-qubit Heisenberg XYZ model, where we employed a complete static anisotropic coupling between the two qubits in presence of an external sinusoidal time-dependent non-uniform magnetic field \cite{coupled}. Motivated by one of the most interesting proposals to create a fundamental quantum computing gate by applying a time-dependent exchange interaction between two quantum dots \cite{solid} , we studied the dynamics of two-coupled two-level atoms (can be considered as two quantum dots) represented by Heisenberg XYZ model. The interaction between the atoms was considered isotropic time-dependent coupling and the entire system is under the effect of an external non-uniform static magnetic field \cite{dynamics}.

In this paper we study the dynamics of entanglement of a two qubit Heisenberg XYZ model for which we employ a complete anisotropic coupling between the two qubits in presence of an external non-uniform magnetic field. We apply both time-dependent magnetic field and time-dependent Heisenberg coupling and present two exact solutions corresponding to two particular integrability conditions possessed by the system. We demonstrate how the dynamics of the system and its entanglement evolution can be controlled using the different parameters of the time-dependent magnetic field and Heisenberg interaction. The discrete symmetries possessed by the system and its effect on the dynamics in the two different subspaces is illustrated. Also we applied two different approximation methods namely the rotated wave and perturbation and compared their results and their range of validity for the special cases of the time-independence of either the magnetic field or the exchange interaction.

This paper is organized as follows. In Sec. II we introduce our model and the underlying symmetries of the system. In Sec. III we discuss the time evolution and the two integrability conditions of the system and the approximation approaches. In Sec. IV we present the entanglement formula in the cases of integrability and approximations and discuss the results. We close with our conclusions in Sec V.
\section{The model and its underlying symmetry}
We consider two coupled qubits through time dependent anisotropic Heisenberg XYZ interaction in presence of a time dependent non-uniform magnetic field applied in the z-direction. The Hamiltonian of the system is given by
\be
\hat{H}(t) = \la_x(t) S_{1x} S_{2x} + \la_y(t) S_{1y} S_{2y} +\la_z(t) S_{1z} S_{2z} +  \om_1(t) S_{1z} +  \om_2(t) S_{2z},
\label{h1}
\ee
where $S_{ij}\;(i=1,2; \; j = x,y,z)$ are the spin half-operators, while $\la_x(t)$, $\la_y(t)$ and $\la_z(t)$ are the time dependent strengths of the Heisenberg interactions in the x, y and z directions respectively. $\om_1(t)$ and $\om_2(t)$ are the external time-dependent magnetic fields. The Hamiltonian can be written in a more convenient form in terms of the operators $S_{i\pm}=S_{ix}\pm\,i\,S_{iy}$ and $S_{z}^{\pm}=S_{1z} \pm S_{2z}$. As a result, the Hamiltonian acquires the form
\bea
\hat{H}(t)= &&\om_+(t) S_{z}^{+} + \om_-(t) S_{z}^{-} + \la_{z}(t)\, S_1\cdot S_2 \nn\\
 && + \left(\la_p(t) - {\la_z(t)\over 2}\right) \left(S_{1+} S_{2-}+ S_{1-} S_{2+}\right) + \la_m(t) \left(S_{1+} S_{2+}+ S_{1-} S_{2-}\right)\;,
\label{h2}
\eea
where $\la_p(t) = (\la_x(t)+\la_y(t))/4$, $\la_m(t) =(\la_x(t)-\la_y(t))/4$, $\om_{\pm}(t)= (\om_1(t) \pm \om_2(t))/ 2$ and $S_1\cdot S_2 = S_{1x}\,S_{2x} + S_{1y}\,S_{2y} + S_{1z}\,S_{2z}$.\\

It is convenient to study the matrix representation of the Hamiltonian of this system in the uncoupled basis, namely $\{|++\rangle, |--\rangle, |+-\rangle, |-+\rangle \}$, where
in terms of these states the Hamiltonian reads
\bea
H=\left[\begin{array}{cccc}
\om_+(t)+{\la_z(t)\over 4} & \la_m(t) & 0 & 0   \\
\la_m(t) & -\om_+(t)+{\la_z(t)\over 4} & 0 & 0  \\
0 & 0 & \om_-(t) -{\la_z(t)\over 4} & \la_p(t)  \\
0 & 0 & \la_p(t) & -\om_-(t) -{\la_z(t)\over 4}
\end{array} \right].
\label{h3}
\eea

In fact many of the characteristic features and properties possessed  by $H$ can be understood and attributed to the role of discrete symmetries enjoyed by the Hamiltonian. It turns out that there are three discrete symmetries plying important roles in understanding the dynamics generated by $H$.

The first discrete symmetry is found to be,
 \be
 S_{ix}\rightarrow -S_{ix},\,\mbox{or}\, S_{iy}\rightarrow -S_{iy}.
  \label{sym1}
  \ee
 The  block diagonal structure of the Hamiltonian matrix can be understood as a consequence of this discrete symmetry. The symmetry defined in eq.~(\ref{sym1}) keeps the Hamiltonian in eq.~(\ref{h1}) invariant and then the spin states can be classified according to their parities under spin reflection of $x$ or $y$ components. It turns out that the states of positive parity are $|++\rangle$ and $|--\rangle$ while those having negative parity are $|+-\rangle$ and $|-+\rangle$. The parity is conserved as a consequence of the symmetry and thus the states of opposite parities never get mixed under time evolution generated by $H$. As a result, the spin state space of the system is splitted into two subspaces and the Hamiltonian is block diagonal as explicitly shown in eq.~(\ref{h3}). Each block of $H$ is controlled by different independent set of parameters except $\la_z(t)$, which is present in both blocks.  The eigenstates of the system and the corresponding energy spectrum can be readily obtained by diagonalizing the two blocks in the matrix $H$. The first subspace is spanned by
\bea
\label{sub1eig}
|\phi_1(t)\rangle & =& \cos{\th_{1}(t)}\; |++\rangle + \sin{\th_{1}(t)}\; |--\rangle,\nn\\
|\phi_2(t)\rangle &=& -\sin{\th_{1}(t)}\; |++\rangle + \cos{\th_{1}(t)}\; |--\rangle,
\eea
where the angle $\th_1(t)$ is defined as
\be
\tan{\left[2\,\th_1(t)\right]}={\la_{m}(t)\over \om_+(t)}.
\label{angth1}
\ee
The associated energy eigenvalues are
\bea
\epsilon_1(t) = {\la_z(t)\over 4} + \eta(t), \; \; \; \; \;
\epsilon_2(t)  = {\la_z(t)\over 4} - \eta(t) ,
\label{eig1}
\eea
where
\be
\eta(t) = \sqrt{\om_+(t)^2 + \la_m(t)^2}.
\label{defeta}
\ee
As to the second subspace, the eigenstates are
\bea
\label{sub2eig}
|\phi_3(t)\rangle &=& \cos{\theta_{2}(t)}\; |+-\rangle + \sin{\theta_{2}(t)}\; |-+\rangle,\nn\\
|\phi_4(t)\rangle &=& -\sin{\theta_{2}(t)}\; |+-\rangle + \cos{\theta_{2}(t)}\; |-+\rangle,
\eea
with the corresponding energy eigenvalues
\bea
\epsilon_3(t) = -{\la_z(t)\over 4} +\zeta(t), \; \; \; \; \;
\epsilon_4(t) = -{\la_z(t)\over 4} - \zeta(t),
\label{eig2}
\eea
where
\be
\tan{\left[2\,\th_2(t)\right]}={\la_{p}(t)\over \om_-(t)}, \;\;\; \; \;
\zeta(t) = \sqrt{\om_-(t)^2 + \la_p(t)^2}.
\label{ang2th}
\ee

As to the second discrete symmetries possessed by $H$, it can be identified as,
\bea
S_{2\a} \rightarrow -S_{2\a}, && \la_x \rightarrow -\la_x,\nn \\
\om_+ \rightarrow \om_-, && \la_z \rightarrow -\la_z \; ,
\label{sym2}
\eea
where $\a \equiv x, y, z$. This symmetry manifests its effect through the spin reflection of $\bf{S}_2$, which maps the two disjoint subspaces in eq.~(\ref{sub1eig}) and eq.~(\ref{sub2eig}) into each other, explicitly as,
\bea
|++\rangle \leftrightarrow  |+-\rangle, & & |--\rangle \leftrightarrow  |-+\rangle,
\label{flip1}
\eea
and consequently all dynamics of the time evolution contained in the subspace spanned by $|+-\rangle$ and $|-+\rangle$ can be obtained from those of $|++\rangle$ and $|--\rangle$  by just the following replacement,
\bea
\la_m \leftrightarrow \la_p, & \la_z \leftrightarrow - \la_z, & \om_+ \leftrightarrow \om_-.
\label{rep1}
\eea
It is clear that the two block in the Hamiltonian given in eq.~(\ref{h3}) can be obtained from each other by the replacement given
in eq.~(\ref{rep1}). The manifestation of the symmetry described in eq.~(\ref{sym2}) would be obscured  having used as a basis the states of definite total spin and total z-component $|++\rangle, |--\rangle, \frac{1}{\sqrt{2}}\left(|+-\rangle + |-+\rangle\right)$ and $\frac{1}{\sqrt{2}}\left(|+-\rangle - |-+\rangle\right)$, and the Hamiltonian matrix turns out to be,
\bea
H=\left[\begin{array}{cccc}
\om_+(t)+{\la_z(t)\over 4} & \la_m(t) & 0 & 0   \\
\la_m(t) & -\om_+(t)+{\la_z(t)\over 4} & 0 & 0  \\
0 & 0 & \la_p(t)-{\la_z(t)\over 4} & \om_-(t)  \\
0 & 0 & \om_-(t) & -\la_p(t)-{\la_z(t)\over 4}
\end{array} \right].
\label{h4}
\eea
Of course, the last symmetry can be equivalently achieved by the reflection of $S_{1\a}$.
Regarding the third discrete symmetry, which is rather obvious and amounts to be reflecting all spins and inverting the sign of $\om_+$ and $\om_-$. The symmetry operation can be realized as,
\bea
\mathbf{S}_1 \rightarrow -\mathbf{S}_1, && \mathbf{S}_2 \rightarrow -\mathbf{S}_2, \nn\\
\om_+ \rightarrow - \om_+, && \om_- \rightarrow - \om_-,
\label{sym3}
\eea
and the spin states are affected accordingly to,
\bea
|++\rangle \leftrightarrow  |--\rangle, & & |+-\rangle \leftrightarrow  |-+\rangle.
\label{flip2}
\eea
As a result of this symmetry all the dynamics contained in the time evolution of the state $|++\rangle$ can be extracted from those of $|--\rangle$ and vice versa. The same applies to the two states $|+-\rangle$ and $|-+\rangle$. The three symmetries described above in eqs.~(\ref{sym1}), (\ref{sym2}) and (\ref{sym3}) turn out to be very useful in checking the correctness of the calculations and also in reducing the calculations labour.

It is important to realize that the eigen states in eq.~(\ref{sub1eig}) and eq.~(\ref{sub2eig}) are not generally stationary states and their associated energy eigenvalues are time dependent as shown in eq.~(\ref{eig1}) and eq.~(\ref{eig2}). These findings seem natural in accord with the explicit time dependent Hamiltonian. The time independent Hamiltonian case is more simple to be analysed, since in this case the eigenstate
are stationary. Further more, the time evolution is a straight forward to be handled. The ground state of the system can be unambiguously determined according to the value of the relevant parameters in eqs.~(\ref{eig1}) and (\ref{eig2}). When $\eta > {\la_z \over 4} + \zeta$, the ground state turns out to be $|\phi_2\rangle$.
On the other hand, for $\eta < {\la_z \over 4} + \zeta$ the ground state is $|\phi_4\rangle$. Also it turns out to be degenerate when $\eta = {\la_z \over 4} + \zeta$ and it can be unambiguously chosen
to be ${1\over\sqrt{2}}\left[e^{i\alpha}\,| \phi_2\rangle + |\phi_4\rangle\right]$. The choice is based on maximum entropy principle that dictates equal mixture between the two degenerate states but leaving arbitrary relative free phase $\alpha$ undetermined. These results are in agreement with those found in \cite{Isotropic_XY}.

\section{Time evolution and integrability conditions}
The time evolution of time dependent Hamiltonian is an intricate problem. Although the time evolution can be numerically solved for any generic Hamiltonian, however exact or approximate analytic solutions are interesting to find. Analytic solutions are important because they can provide a wealth of information about the feature of solutions with minimal numerical works, even more they could serve as a checking benchmark for the validity of numerical solutions.

The time evolution of any state ket vector is governed by Schrodinger equation in the form,
\be
H\,| \Psi\rangle = i {\partial | \Psi\rangle \over \partial t}.
\ee
Knowing the time evolution of the eigen states enables us to determine the time evolution of any arbitrary state. Starting with time evolution in the subspace spanned by the two eigen states $|\phi_1(t)\rangle$ and $|\phi_2(t)\rangle$ as defined in eq.~(\ref{sub1eig}). Assuming an initial state, at $t=0$, in the form,
\be
 |\phi_1(0)\rangle=\cos \th_{10} |++\rangle + \sin \th_{10} |--\rangle,
 \label{in1}
 \ee
  where
$\th_{10}=\th_1(0)$. The initial state specified in eq.~(\ref{in1}) evolves in time and at later time $t$  becomes $|\psi_1(t)\rangle$, which is described by,
\be
|\psi_1(t)\rangle = g_1(t)\, \cos \th_{10} |++\rangle + g_2(t)\, \sin \th_{10} |--\rangle \;,
\ee
where $g_1(t)$ and $g_2(t)$ are unknown functions of $t$ to be specified through Schrodinger equation
supplemented by the initial condition $g_1(0)=1 = g_2(0)=1$. The time evolution implied by Schrodinger equation for the functions $g_1(t)$ and $g_2(t)$ can be  expressed as first order coupled differential equations. The resulting equations can be written in a compact matrix form as,
\be
\dot{x} = - i\, H_I x,
\label{evol1}
\ee
where the dot denotes the derivative with respect to time, while $x$ and $H_I$ are,
\be
\begin{array}{ll}
\begin{array}{lll}
x & = &
\left[
\begin{array}{l}
g_1(t)\, \cos \th_{10}\\
g_2(t)\, \sin \th_{10}
\end{array}
\right],
\end{array}
&
\begin{array}{lll}
H_I &=&
\left[
\begin{array}{cc}
\om_+(t)+{\la_z(t)\over 4} & \la_m(t)   \\
\la_m(t) & -\om_+(t)+{\la_z(t)\over 4}
\end{array}
\right],
\end{array}
\end{array}
\label{evol2}
\ee
the matrix $H_I$, as expected, is the Hamiltonian matrix in the subspace generated by $|\phi_1(t)\rangle$ and $|\phi_2(t)\rangle$.

A more illuminating form for eq.~(\ref{evol1}) can be reached by transforming the variable $x$ into a new one $u$ through $u = S_I\,x$, where the matrix $S_I$ is given by
\be
\begin{array}{lll}
S_I & = &
\left[
\begin{array}{ll}
\cos{\th_1(t)} & \sin{\th_1(t)}\\
-\sin{\th_1(t)} & \cos{\th_1(t)}
\end{array}
\right],
\end{array}
\ee
where the angle $\th_1(t)$ is defined in eq.~(\ref{angth1}) and the matrix $S_I$ diagonalize $H_I$. After some simple algebraic manipulations, the resulting equation governing $u$ can be casted in the
form,
\be
\dot{u}+ i\,\left(- \,\dot{\th}_1(t)\,\sig_2 + {\la_z(t)\over 4}+\eta(t)\, \sig_3\right)u=0,
\label{equ}
\ee
where $\sig_2$ and $\sig_3$ are $2\times 2$ Pauli spin matrices given respectively as,
\be
\begin{array}{ll}
\begin{array}{lll}
\sig_2 & = &
\left[
\begin{array}{ll}
0 & -i \\
i & 0
\end{array}
\right],
\end{array}
&
\begin{array}{lll}
\sig_3 & = &
\left[
\begin{array}{ll}
1 & 0 \\
0 & -1
\end{array}
\right].
\end{array}
\end{array}
\label{pauli23}
\ee
The factor ${\la_z(t)/ 4}$ should be understood as multiplied by $2 \times 2$ identity matrix. The differential equation in eq.~(\ref{equ}) is subjected to the initial condition,
\be
\begin{array}{lll}
u(0) &=&
\left[
\begin{array}{l}
1\\
0
\end{array}
\right].
\end{array}
\label{inu}
\ee
It is worthy to mention some explanation for the used notations. The subscript index I reveals that the quantity is restricted to the space spanned by first two eigenstates namely $|\phi_1(t)\rangle$ and $|\phi_2(t)\rangle$, while using subscript index II would mean restriction to subspace spanned by $|\phi_3(t)\rangle$ and $|\phi_4(t)\rangle$. The x's, y's, z's and w's coefficients in investigating time evolution refer to ones starting with initial states $|\phi_1(0)\rangle$, $|\phi_2(0)\rangle$, $|\phi_3(0)\rangle$ and $|\phi_4(0)\rangle$  respectively.
\subsection{The first integrability condition}
A straight forward integrability condition enabling closed form solutions for eq.~(\ref{equ}) is given by,
\be
\dot{\th}_1(t)=0,
\label{sintg1}
\ee
and the obtained solutions in this case take the form,
\be
\begin{array}{lll}
\begin{array}{lll}
u(t)& =&
\left[
\begin{array}{l}
I_1 \\\\
0
\end{array}
\right]
\end{array}
&\Rightarrow&
\begin{array}{lll}
x(t) &=&
\left[
\begin{array}{l}
\cos{\th_{10}}\,I_1 \\\\
\sin{\th_{10}}\,I_1
\end{array}
\right],
\end{array}
\end{array}
\label{intg1}
\ee
the integrability condition in eq.~(\ref{sintg1}) is fulfilled when $\om_+(t) \propto \la_m(t)$. Similarly, the initial state $|\phi_2(0)\rangle$ under time evolution is described by $y$ that amounts to
\be
\begin{array}{lll}
y(t) &=&
\left[
\begin{array}{l}
-\sin{\th_{10}}\,I_2 \\\\
\cos{\th_{10}}\,I_2
\end{array}
\right].
\end{array}
\label{intg2}
\ee
The corresponding $z$ and $w$ are easily determined to be
\be
\begin{array}{lll}
\begin{array}{lll}
z(t)& =&
\left[
\begin{array}{l}
\cos{\th_{20}}\,I_3 \\\\
\sin{\th_{20}}\,I_3
\end{array}
\right],
\end{array}
& &
\begin{array}{lll}
w(t) &=&
\left[
\begin{array}{l}
-\sin{\th_{20}}\,I_4 \\\\
\cos{\th_{20}}\,I_4
\end{array}
\right].
\end{array}
\end{array}
\label{intg3}
\ee
where
\be
I_{j}=\exp{\left(-i\,\int_0^t \epsilon_j(t')\,dt'\right)}
\ee
\subsection{The second integrability condition}
Another simple integrability condition is,
\be
\dot{\th}_1(t) = \kappa\; \eta(t) \Rightarrow \dot{\th}_1(t) = \kappa\; {\la_m(t)\over \sin{[2\,\th_1(t)]}},\; \kappa \neq 0,
\label{sintg2}
\ee
where $\kappa$ is constant. The condition upon integration leads to,
\be
\cos{[2\, \th_1(t)]} = -2 \,\kappa \,\int_0^t \la_m(t')\, dt' + \cos{(2\, \th_{10})},
\label{cos1}
\ee
which in turn implies
\be
\tan{[2\,\th_1(t)]}= {\sqrt{1- \left(-2\, \kappa\, \int_0^t \la_m(t')\, dt' + \cos{(2\,\th_{10})}\right)^2}\over
\left(-2\, \kappa \,\int_0^t \la_m(t')\, dt' + \cos{(2\, \th_{10})}\right)}.
\ee
In this case the solutions for eq.~(\ref{equ}) are,
\be
u(t) = \exp{\left\{i\, \int_0^t\,\left[ \dot{\th}_1(t')\left( \sig_2 - \kappa ^{-1}\, \sig_3\right) - {\la_z(t')\over 4}\right]\,dt'\right\}}\, u(0) \;,
\label{eqic21}
\ee
which upon using eq.~(\ref{sintg2}) becomes
\be
u(t) = \Lambda \times \left( \cos{\delta_{\th_1(t)}} -i\,\sig_3\,{\kappa^{-1} \over \sqrt{1+\kappa^{-2}}}\,\sin{\delta_{\th_1(t)}}
+ i\,\sig_2 {\sin{\delta_{\th_1(t)}}\over \sqrt{1+\kappa^{-2}}} \right)\, u(0) \; ,
\label{eqic22}
\ee
where,
\be
\delta_{\th_1(t)}= \left(\th_1(t) - \th_{10}\right)\,\sqrt{1+\kappa^{-2}}\;, \; \; \;
\Lambda = \exp{\left(-i\,\int_0^t {\la_z(t')\over 4}\,dt'\right)}\;.
\label{defdelta}
\ee
The corresponding column vector $x(t)$ in terms of its components, $x_1(t)$ and $x_2(t)$, is given as,
\bea
\nonumber x_1(t) &=&
   \Lambda \,\times\bigg[\cos{\th_{1}(t)}\,\left( \cos{\delta_{\th_1(t)}}
-i\,\kappa^{-1}\,\bar{\kappa} \,\sin{\delta_{\th_1(t)}}\right)\\
\nonumber&+& \sin{\th_{1}(t)}\,\bar{\kappa}\,\sin{\delta_{\th_1(t)}}\bigg],\nn\\
\nonumber x_2(t) &=&
  \Lambda \,\times\bigg[\sin{\th_{1}(t)}\,\left( \cos{\delta_{\th_1(t)}}
-i\,\kappa^{-1}\,\bar{\kappa}\,\sin{\delta_{\th_1(t)}}\right)\\
&-& \cos{\th_{1}(t)}\,\bar{\kappa}\,\sin{\delta_{\th_1(t)}}\bigg]
\label{eqic23}.
\eea
Similarly, the initial state $|\phi_2(0)\rangle$ evolves in time according to,
\bea
\nonumber y_1(t) &=&
  \Lambda \,\times\bigg[-\sin{\th_{1}(t)}\,\left( \cos{\delta_{\th_1(t)}}
+i\,\kappa^{-1}\,\bar{\kappa} \,\sin{\delta_{\th_1(t)}}\right)\\
\nonumber&+& \cos{\th_{1}(t)}\,\bar{\kappa}\,\sin{\delta_{\th_1(t)}}\bigg],\nn\\
\nonumber y_2(t) &=&
  \Lambda \,\times\bigg[\cos{\th_{1}(t)}\,\left( \cos{\delta_{\th_1(t)}}
+i\,\kappa^{-1}\, \bar{\kappa}\,\sin{\delta_{\th_1(t)}}\right)\\
&+& \sin{\th_{1}(t)}\,\bar{\kappa}\,\sin{\delta_{\th_1(t)}}\bigg]
\label{eqic24}.
\eea

where,
\be
\bar{\kappa}={1 \over \sqrt{1+\kappa^{-2}}}
\label{kappabar}.
\ee
 In the second subspace generated by $|\phi_3(t)\rangle$ and $|\phi_4(t)\rangle$, the corresponding integrability condition is
\be
\dot{\th}_2(t) = \chi\; \zeta(t) \Rightarrow \dot{\th}_2(t) = \chi\; {\la_p(t)\over \sin{[2\,\th_2(t)]}}
\label{s2intg2}
\ee
where $\chi$ is constant. The condition upon integration leads to,
\be
\cos{[2\, \th_2(t)]} = -2 \,\chi \,\int_0^t \la_p(t')\, dt' + \cos{(2\, \th_{20})}.
\ee
In a similar way the solutions for time evolution can be obtained as those of eqs.~(\ref{eqic23}) and (\ref{eqic24}), which come out to be
\bea
\nonumber z_1(t)&=&
 \bar{\Lambda}\,\times\bigg[\cos{\th_{2}(t)}\,\left( \cos{\delta_{\th_2(t)}}
-i\,\chi^{-1}\,\bar{\chi}\,\sin{\delta_{\th_2(t)}}\right)\\
 \nonumber&+&\sin{\th_{2}(t)}\,\bar{\chi}\,\sin{\delta_{\th_2(t)}}\bigg],\nn\\
\nonumber z_2(t) &=&
 \bar{\Lambda} \,\times\bigg[\sin{\th_{2}(t)}\,\left( \cos{\delta_{\th_2(t)}}
-i\,\chi^{-1}\,\bar{\chi}\,\sin{\delta_{\th_2(t)}}\right)\\
\nonumber &-&\cos{\th_{2}(t)}\,\bar{\chi}\,\sin{\delta_{\th_2(t)}}\bigg],\nn\\
\nonumber w_1(t) &=&
\bar{\Lambda} \,\times\bigg[-\sin{\th_{2}(t)}\,\left( \cos{\delta_{\th_2(t)}}
+i\,\chi^{-1}\,\bar{\chi}\,\sin{\delta_{\th_2(t)}}\right)\\
\nonumber &+& \cos{\th_{2}(t)}\,\bar{\chi}\,\sin{\delta_{\th_2(t)}}\bigg],\nn\\
\nonumber w_2(t) &=&
 \bar{\Lambda} \,\times\bigg[\cos{\th_{2}(t)}\,\left( \cos{\delta_{\th_2(t)}}
+i\,\chi^{-1}\,\bar{\chi}\,\sin{\delta_{\th_2(t)}}\right)\\
 &+& \sin{\th_{2}(t)}\,\bar{\chi}\,\sin{\delta_{\th_2(t)}}\bigg]
\label{eqic25}.
\eea
where $\delta_{\th_2(t)}$ is defined as
\be
\delta_{\th_2(t)}= \left(\th_2(t) - \th_{20}\right)\,\sqrt{1+\chi^{-2}}.
\label{defdelta2}
\ee
and
\be
\bar{\chi} = {1 \over \sqrt{1+\chi^{-2}}}\;, \; \; \;
\bar{\Lambda} = \exp{\left(i\,\int_0^t {\la_z(t')\over 4}\,dt'\right)}\;.
\ee

Equation~(\ref{cos1}) adds an extra restriction to second integrability condition, due to the constrain $\left|\cos(2 \theta_1(t))\right| \leq 1$. Considering the Heisenberg interactions difference $\la_m(t)$ to be a time-dependent function of the form
\be
\la_m(t) = \mu_m\,\sin{(\beta_{m} \, t +\phi_{m})} \; ,
\ee
where $\mu_m$, $\beta_{m}$ and $\phi_{m}$ are the amplitude, frequency and initial phase of $\lambda_{m}(t)$, respectively. The restriction in the second integrability case manifests itself through the relation among the system parameters. In the case of $\th_{10}\neq 0$, the condition takes the form
\be
\frac{\beta_{m}}{2\,\kappa\,\mu_m}\geq \mbox{max}\left(\frac{2}{1+\cos{(2\, \th_{10})}},\frac{2}{1-\cos{(2\, \th_{10})}}\right),
\ee
while in the case of $\th_{10}=0$, it becomes
\be
\phi_{m}=0,\quad\quad\quad
\left|\frac{4\,\kappa\,\mu_m}{\beta_{m}}\right| \; \leq \; 1.
\ee
\subsection{Perturbation versus rotated wave approximations}
In this section we apply and compare two of the widely used methods of approximations namely the perturbation theory and the rotated wave approximation (RWA). We seek to explore their range of validity in treating our system and compare their results.
We start with the perturbation theory, where we consider eqs.~(\ref{evol1}) and (\ref{evol2}) and assume that $\om_+$ and $\la_z$ are constants where $\la_m$ is small to be a considered as perturbation. Therefore, $H_I$ represents two levels system where
\bea
E_1 = -\om_+ + {\la_z \over 4},\;\;\;\;\;
E_2 = \om_+ + {\la_z \over 4}.
\eea
The difference between the two levels is given by $E_{2}-E_{1}=2\om_+$.
The evolution due to $\la_z$ can be factored out, so $H_I$ takes the form
\be
H_I =
\left[
\begin{array}{ll}
\om_+ & \la_m(t)   \\
\la_m(t) & -\om_+
\end{array}
\right].
\ee
Then the time evolution is controlled by a system of two coupled differential equations which can be written in a matrix for as,
\bea
\label{pert}
\left[\begin{array}{c}
\dot{{x}_1} \\
\dot{{x}_2}
\end{array} \right] = -i \;
\left[\begin{array}{cc}
\om_+ &  \la_{m}(t)   \\
\la_{m}(t) & -  \om_+
\end{array} \right] \;
\left[\begin{array}{c}
{x}_1   \\
{x}_2
\end{array} \right],
\eea
Assuming the system is initially in the  state $\left|++\right\rangle$ and suffering a weak transition due to the small perturbation $\la_m(t)$. The weak transition demands that $\left|x_1\right| \gg \left|x_2\right|$ and $\left|x_1\right|\approx 1$ to be fulfilled.

Employing perturbation technique to solve the system in  Eq.(\ref{pert}), taking into account that $\la_{m}(t)$ is small, one obtains the following solution
\bea
\label{x12}
x_1 &\approx& e^{-i\,\om_{+}\, t},\nn\\
x_2 &\approx& -i\, e^{i\,\om_{+}\,t}\,\int\limits_{0}^{t} \la_{m}(\tau)\, e^{-2i\,\om_{+}\,\tau}\,d\tau,
\eea
which is clearly consistent to represent a weak transition.

To be more specific we  consider a weak harmonic perturbation in the form,
\be
\la_m(t) = \mu_m\,\sin{\left(\beta_m \, t\right)} \;,
\ee
where $\mu_m$ is the amplitude and $\beta_m$ is the frequency of $\la_m$.  The solution $x_2(t)$ in eq.(\ref{x12}) becomes,
\be
x_2=\frac{i}{2}\,e^{i\,\om_{+}\,t}\,\mu_m\,
\left[\frac{e^{i\,(\beta_m-2\,\om_+)\,t}}{(\beta_m-2\,\om_+)}+
\frac{e^{-i\,(\beta_m+2\,\om_+)\,t}}{(\beta_m+2\,\om_+)}-
\frac{1}{(\beta_m-2\,\om_+)}-\frac{1}{(\beta_m+2\,\om_+)}\right] \;.
\ee
Considering the near resonance case ($\beta_m\approx 2\,\om_+$), and after keeping only the  dominant terms
proportional to ${1\over \beta_m -2\,\om_+}$, we get
\bea
|x_2|^2\approx{\mu_m^2 \over{(\beta_m-2\,\om_+)^2}}\,\sin^2{((\beta_m-2\,\om_+)\,t/2)}.
\label{solnrot10}
\eea
It is important to emphasize that consistency with perturbation requires
\be
\left|{\mu_m \over \left(\beta_m-2\,\om_+\right)}\right| \ll 1 .
\ee

Now let us turn our attention to the rotated wave approximation, which is widely used in quantum optics and magnetic resonance. In this approximation, terms in the system Hamiltonian that oscillate rapidly are neglected \cite{RWA}. The time evolution in eqs.~(\ref{evol1}) and (\ref{evol2}) can be rewritten as
\be
\dot{x'} = - i\, H'_I x',
\label{mateqR1}
\ee
where $x'$ and $H'_I$ are,
\be
\begin{array}{ll}
\begin{array}{lll}
x' & = &
\left[
\begin{array}{l}
\bar{\Lambda}\,x_1\\
\bar{\Lambda}\,x_2
\end{array}
\right],
\end{array}
&
\begin{array}{lll}
H'_I &=&
\left[
\begin{array}{ll}
\om_+(t) & \la_m(t)   \\
\la_m(t) & -\om_+(t)
\end{array}
\right],
\end{array}
\end{array}
\ee
Considering a constant $\om_+$, while assuming a sinusoidal time varying $\la_m(t)$ in the form
\be
\la_m(t) = \mu_m\,\sin{\left(\beta_m \, t +\phi_m\right)}
\ee
where $\mu_m$ is the amplitude, $\beta_m$ is the frequency and $\phi_m$ is the initial phase of $\la_m$.
Rotating the coordinates of the system ($x'_1 \rightarrow e^{i\beta_m t} x'_1$, $x'_2 \rightarrow x'_2$) and then dropping the high frequency terms, we obtain
\bea
\label{roteq1}
\left[\begin{array}{c}
\dot{\tilde{x}}_1 \\
\dot{\tilde{x}}_2
\end{array} \right] = -i \;
\left[\begin{array}{cc}
-\beta_m+\om_+ & \;\;\;i\, {\mu_m\over 2}\, e^{-i\phi_m}   \\
-i\, {\mu_m\over 2}\, e^{i\phi_m} & -  \om_+
\end{array} \right] \;
\left[\begin{array}{c}
\tilde{x}_1   \\
\tilde{x}_2
\end{array} \right],
\eea
where $\tilde{x}_1=x'_1 e^{i \beta_m t}$ and $\tilde{x}_2 = x'_2$. Diagonalizing the Hermitian matrix of coefficients in Eq.~(\ref{roteq1}) and after some lengthy algebraic calculations we obtain
\bea
\nonumber x_1 &=& \Lambda\,e^{-i \beta_m t /2}\times \bigg\{ \left[\cos \left(\gamma_m t/2\right) + i \sin\left(\gamma_m t/2\right)\, \cos 2 \theta_m\right] \cos \theta_{10}\\
\nonumber &+& \sin 2 \theta_m \sin (\gamma_m t/2) e^{-i \phi_m} \sin \theta_{10} \bigg\},\nn\\
\nonumber x_2 &=& \Lambda\,e^{i \beta_m t /2} \times \bigg\{-\sin 2 \theta_m \sin \left(\gamma_m t/2\right) e^{i \phi_m} \cos \theta_{10}\\
&+& \left[\cos \left(\gamma_m t/2\right) - i \sin \left(\gamma_m t/2\right) \cos 2 \theta_m\right] \sin \theta_{10}   \bigg\} \;,
\label{solnrot1}
\eea
where $\gamma_m$ and $\theta_m$ are given as,
\bea
\gamma_m &=& \sqrt{\mu_m^{2}+(\beta_m - 2 \om_+)^2},\nn\\
\theta_m &=& \frac{1}{2} \tan^{-1}\left[\mu_m/ (\beta_m - 2 \om_+)\right].
\eea

Alternatively, one can consider a sinusoidal time varying  $\om_+(t)$ in the form
\be
\om_+(t) = \mu_+\,\sin{(\beta_+ \, t +\phi_+)},
\ee
while keeping $\la_m$ as a constant. In this case the solutions are easily found to be:
\bea
x_1 &=& {x^+ + x^- \over \sqrt{2}},\nn \\
x_2 & =& {x^+ - x^- \over \sqrt{2}},
\eea
where $x^+$ and $x^-$ are
\bea
\nonumber x^+ &=& {1\over \sqrt{2}}\,\Lambda\,e^{-i \beta_+ t /2}\times\bigg\{ \left[\cos (\gamma_+ t/2) + i \sin (\gamma_+ t/2) \cos 2 \theta_+\right]\, \left(\cos \theta_{10}+\sin \theta_{10}\right)\\
\nonumber &+& \sin 2 \theta_+ \sin (\gamma_+ t/2) e^{-i \phi_+} \left(\cos \theta_{10}-\sin \theta_{10}\right) \bigg\},\nn\\
\nonumber x^- &=&{1\over \sqrt{2}}\,\Lambda\,e^{i \beta_+ t /2} \times\bigg\{-\sin 2 \theta_+ \sin (\gamma_+ t/2) e^{i \phi_+} \left(\cos \theta_{10}+\sin \theta_{10}\right)\\
&+& \left[\cos (\gamma_+ t/2) - i \sin (\gamma_+ t/2) \cos 2 \theta_+\right] \left(\cos \theta_{10}-\sin \theta_{10}\right)   \bigg\}.
\eea
where $\gamma_+$ and $\theta_+ $ are given as,
\bea
\gamma_+ &=&\sqrt{\mu_+^{2}+(\beta_+ - 2 \la_m)^2},\nn\\
\theta_+ &=& \frac{1}{2} \tan^{-1}[\mu_+/ (\beta_+ - 2 \la_m)].
\label{gamth}
\eea
The time evolution of the second orthogonal state in subspace (I) can be done by applying the replacement $\cos{\th_{10}}\Rightarrow - \sin{\th_{10}}$ and $\sin{\th_{10}}\Rightarrow \cos{\th_{10}}$. The corresponding time evolution in the second subspace (II) can be found by utilizing the symmetry described in eqs.~(\ref{sym2})-(\ref{rep1}). It is worthy to mention that all our results for time evolution (exact or approximate) can be easily shown to be consistent with the constrained imposed by the symmetries as explained in section~2.\\

Now let us compare our obtained results using RWA and perturbation. In order to get faithful comparison, one should set $\theta_{10}=0$  in eq.~(\ref{solnrot1}) and substitute for $\gamma_m$ and $\theta_m$ their values according to eq.(\ref{gamth}).
Thus, we obtain,
\bea
|x_2(t)|^2={{\mu_m^2}\over {\mu_m^{2}+(\beta_m - 2 \om_+)^2}}\,\sin^2\left({\sqrt{\mu_m^{2}+(\beta_m - 2 \om_+)^2}\,t/ 2}\right),
\eea
which for a small $\mu_m$ leads to,
\bea
|x_2(t)|^2\approx{{\mu_m^2}\over {(\beta_m-2\,\om_+)^2}}\,\sin^2\left({(\beta_m-2\,\om_+)\,t/2}\right).
\eea
 This result correspond to what we got in eq.~(\ref{solnrot10}), which shows that the rotating wave approximation goes beyond perturbation result. In fact, by using rotating wave approximation we get rid of the pole at $\beta_m=2\,\om_+$, which is an artefact of perturbation.
\section{Time evolution of entanglement of two qubits}
The amount of entanglement between two quantum system is a monotonic function of what is called the concurrence \cite{woott98}. The concurrence varies from a minimum value of zero to a maximum of one coinciding with the entanglement function range. Therefore, the concurrence itself is considered as a measure of entanglement. To calculate the concurrence one needs to evaluate the matrix
\be
R=\rho(\sigma_y \otimes \sigma_y) \rho^* (\sigma_y \otimes \sigma_y),
\label{rho}
\ee
where $\rho$ is the density matrix of the system evaluated in the coupled representation basis mentioned before and $\rho^*$ is its complex conjugate. The concurrence is defined as
\be
C = \mbox{max}\{\lambda_1-\lambda_2-\lambda_3-\lambda_4, 0\},
\label{condef}
\ee
where $\lambda_1$, $\lambda_2$, $\lambda_3$, $\lambda_4$ are the positive square roots of the eigenvalues of $R$ in the descending order. The evaluation of the concurrence for the pure state
 \be
|\psi\rangle = f_{++}\, |++\rangle\; +\; f_{--}\,|--\rangle\; +\; f_{+-} |+-\rangle\; +\; f_{-+} |-+\rangle,
\label{genpsiu}
\ee
is straightforward, where in this case $\rho=|\psi\rangle \langle\psi|$ and yields
\be
\label{conu}
C =  2\,\left|f_{++}\,f_{--} - f_{+-}\,f_{-+}\right|.
\ee
Having used the state expanded in terms of coupled representation basis as,
\be
|\psi\rangle = f_1 |++\rangle + f_2 |--\rangle + f_3 \frac{1}{\sqrt{2}}\left(|+-\rangle + |-+\rangle\right) + f_4 \frac{1}{\sqrt{2}}\left(|+-\rangle - |-+\rangle\right),
\label{genpsic}
\ee
we get the concurrence as
\be
\label{conc}
C =  \left|2 f_1 f_2 - (f_{3}^{2}-f_{4}^{2})\right|,
\ee
which is, of course, equivalent for the one obtained in eq.~(\ref{conu}) but not
enough transparent to cope with the symmetry explained in eqs.~(\ref{sym2})-(\ref{rep1}).

To reduce the computational labor we can restrict ourself to the subspace (I) in computing the concurrence. The result of the other subspace (II) can be in turn derived just by applying the symmetry elaborated in eqs.~(\ref{sym2})-(\ref{rep1}). In fact, as clear from the formula in eq.~(\ref{conu}), the two subspaces produce interfering contributions to entanglement, that means they can reinforce or cancel each other in the sense of constructive and destructive interference.

Any generic initial state of the system living in the subspace ($I$) can be written as
\bea
\label{psigenIin}
|\psi(0)\rangle &=& a \; |++\rangle\; +\; b \; |--\rangle,\;\; \mbox{and}\; |a|^2 + |b|^2=1,\nn\\
&=& \left( a\; \cos{\th_{10}} + b\; \sin{\th_{10}}\right)\,|\phi_1(0)\rangle
 + \left(- a\; \sin{\th_{10}} + b\; \cos{\th_{10}}\right)\,|\phi_2(0)\rangle,
\eea
then the initial state evolves in time, through $H$, to
\bea
|\psi(t)\rangle &=&  f_{++}\;|++\rangle\ + f_{--}\; |--\rangle.
 \label{psigenIt}
 \eea
It is easy to read the appropriate coefficients $f_{++}$ and $f_{--}$ determining the evolved state to be,
\bea
f_{++} &=& \left( a\; \cos{\th_{10}} + b\; \sin{\th_{10}}\right)\,x_1
+ \left(- a\; \sin{\th_{10}} + b\; \cos{\th_{10}}\right)\,y_1\,\nn \\
f_{--} &=& \left( a\; \cos{\th_{10}} + b\; \sin{\th_{10}}\right)\,x_2
+ \left(- a\; \sin{\th_{10}} + b\; \cos{\th_{10}}\right)\,y_2.
\label{coefI}
\eea
The concurrence in eq.~(\ref{conu}) can be calculated and this leads to,
\bea
C &=& 2\,\left|\left( a\; \cos{\th_{10}} + b\; \sin{\th_{10}}\right)^2 \,x_1\,x_2 +
\left(- a\; \sin{\th_{10}} + b\; \cos{\th_{10}}\right)^2\, y_1\, y_2\right.\,\nn \\
 &&+
\left.\left( a\; \cos{\th_{10}} + b\; \sin{\th_{10}}\right)\,\left(- a\; \sin{\th_{10}} + b\; \cos{\th_{10}}\right)\,\left(x_1\, y_2 + x_2\,y_1\right)\right|.
\label{conpsiI}
\eea
Having a more generic state living in all subspaces like,
\bea
\label{psigen}
|\psi(0)\rangle &=& a \; |++\rangle\; +\; b \; |--\rangle \;+\;c \; |+-\rangle \; +\;
 d \; |-+\rangle,\nn\\
 && \mbox{and}\; |a|^2 + |b|^2 + |c|^2 + |d|^2=1.
 \eea
The concurrence can be straight forwardly  calculated to be,
\bea
C &=& 2\,\left|\left( a\; \cos{\th_{10}} + b\; \sin{\th_{10}}\right)^2 \,x_1\,x_2 +
\left(- a\; \sin{\th_{10}} + b\; \cos{\th_{10}}\right)^2\, y_1\, y_2\right.\,\nn \\
 &&+
\left( a\; \cos{\th_{10}} + b\; \sin{\th_{10}}\right)\,\left(- a\; \sin{\th_{10}} + b\; \cos{\th_{10}}\right)\,\left(x_1\, y_2 + x_2\,y_1\right)\nn \\
&&- \left( c\; \cos{\th_{20}} + d\; \sin{\th_{20}}\right)^2 \,z_1\,z_2 -
\left(- c\; \sin{\th_{20}} + d\; \cos{\th_{20}}\right)^2\, w_1\, w_2\nn \\
&&-\left( c\; \cos{\th_{20}} + d\; \sin{\th_{20}}\right)\,\left(- c\; \sin{\th_{20}} - d\; \cos{\th_{20}}\right)\,\left(z_1\, w_2 + z_2\,w_1\right)\left.\right|.
\label{conpsig}
\eea
Although we can compute the concurrence for time evolved state starting from any generic initial state. It is more instructive to restrict ourself to initial  special states like disentangled or maximally entangled one. More specifically we consider the initial states
$|++\rangle$ and $\frac{1}{\sqrt{2}}\left(|++\rangle + |--\rangle\right)$, and studying their time evolutions and their associated concurrences in the different regime of exact and approximate solution. This is seen in section 3.
\subsection{Entanglement formula for the first integrability condition}
We devote this section to consider different initial states for the systems and investigate the time evolution of them at different parameter values. Applying the first integrability condition we set $\lambda_{m}(t)=k\,\omega_{+}(t)$, where $k$ is a constant.\\
\begin{figure}[htbp]
\begin{minipage}[c]{\textwidth}
 \centering
   \subfigure{\label{fig:1a}\includegraphics[width=5.5cm]{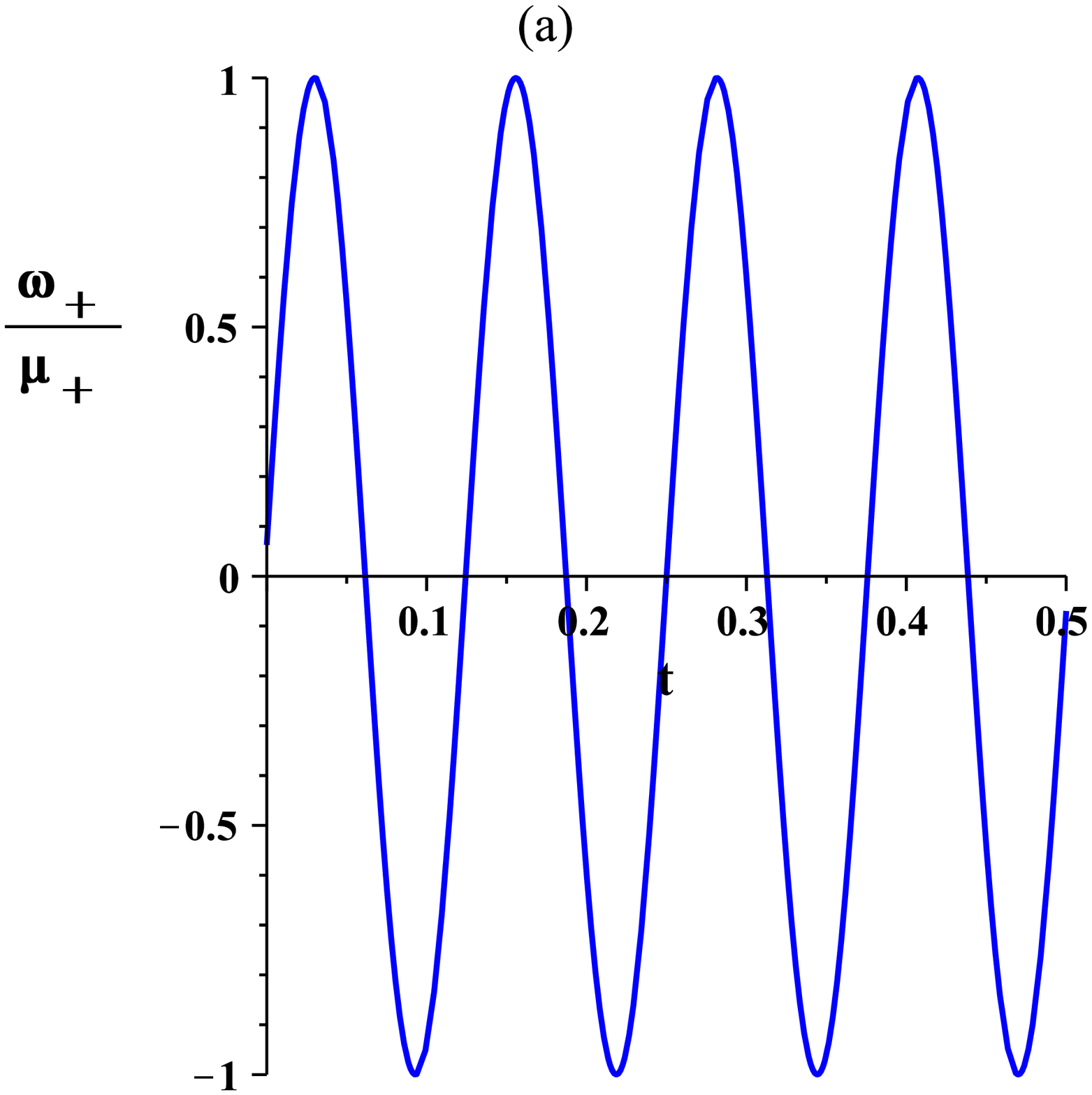}}\quad
   \subfigure{\label{fig:1b}\includegraphics[width=5.5cm]{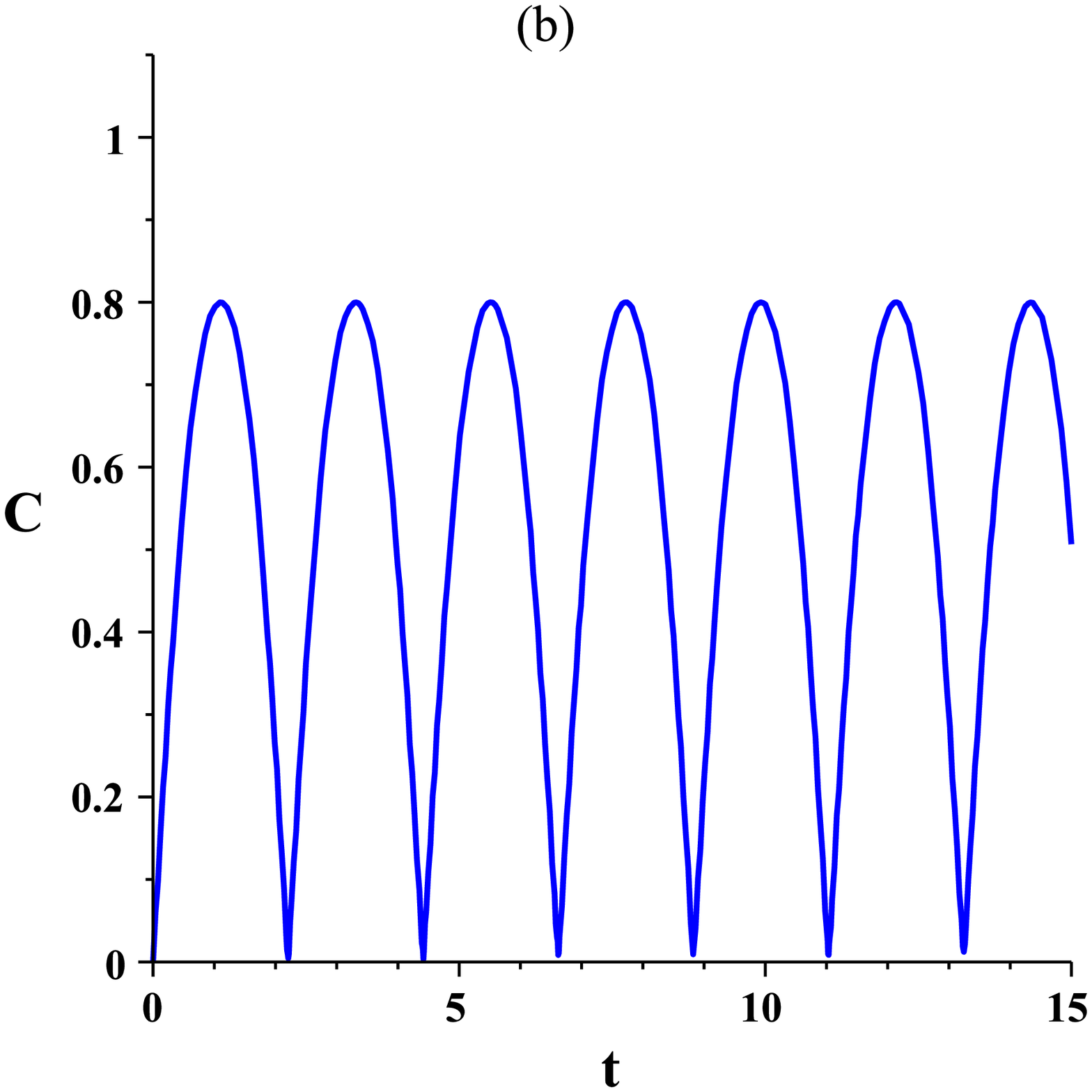}}\\
   \subfigure{\label{fig:1c}\includegraphics[width=5.5cm]{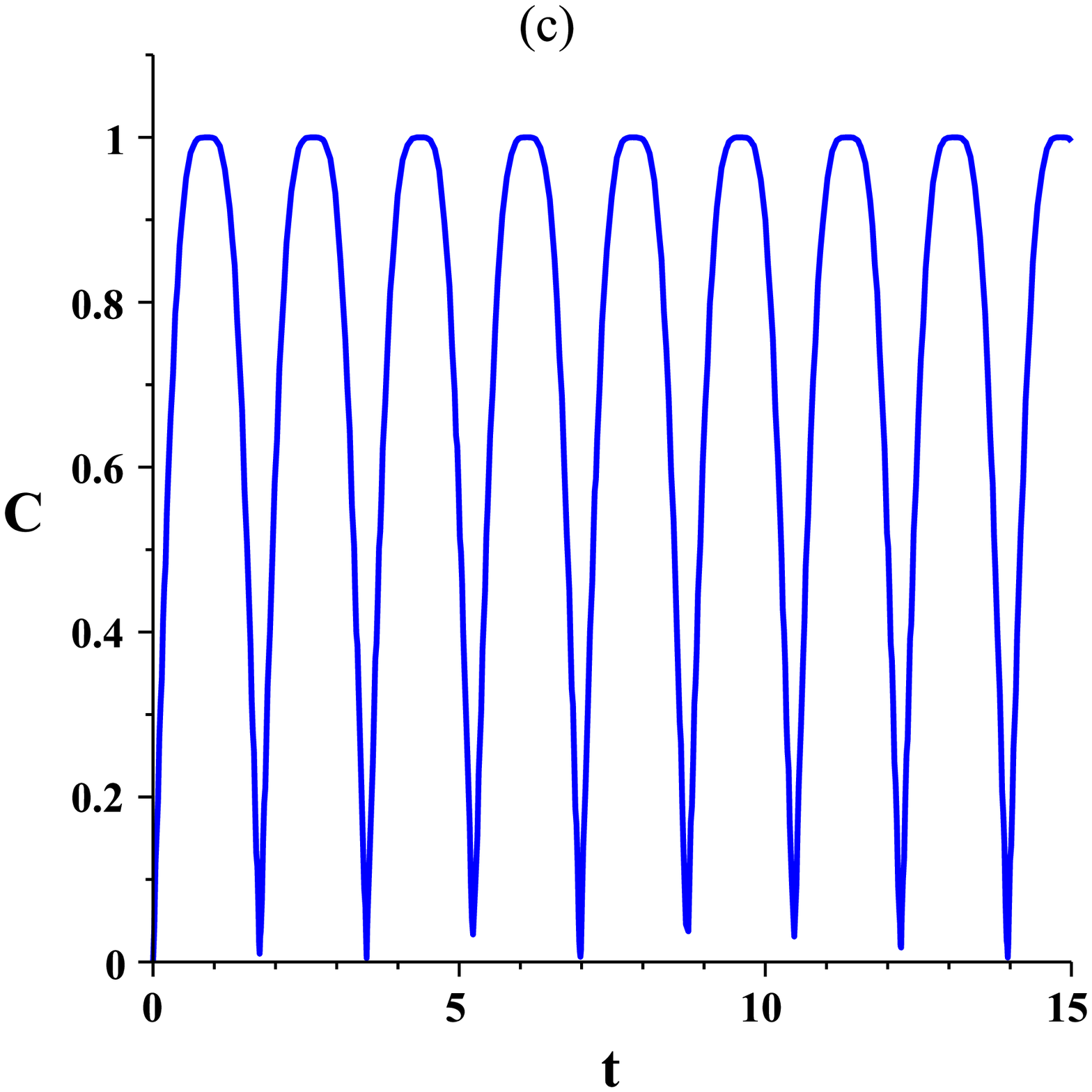}}\quad
   \subfigure{\label{fig:1d}\includegraphics[width=5.5cm]{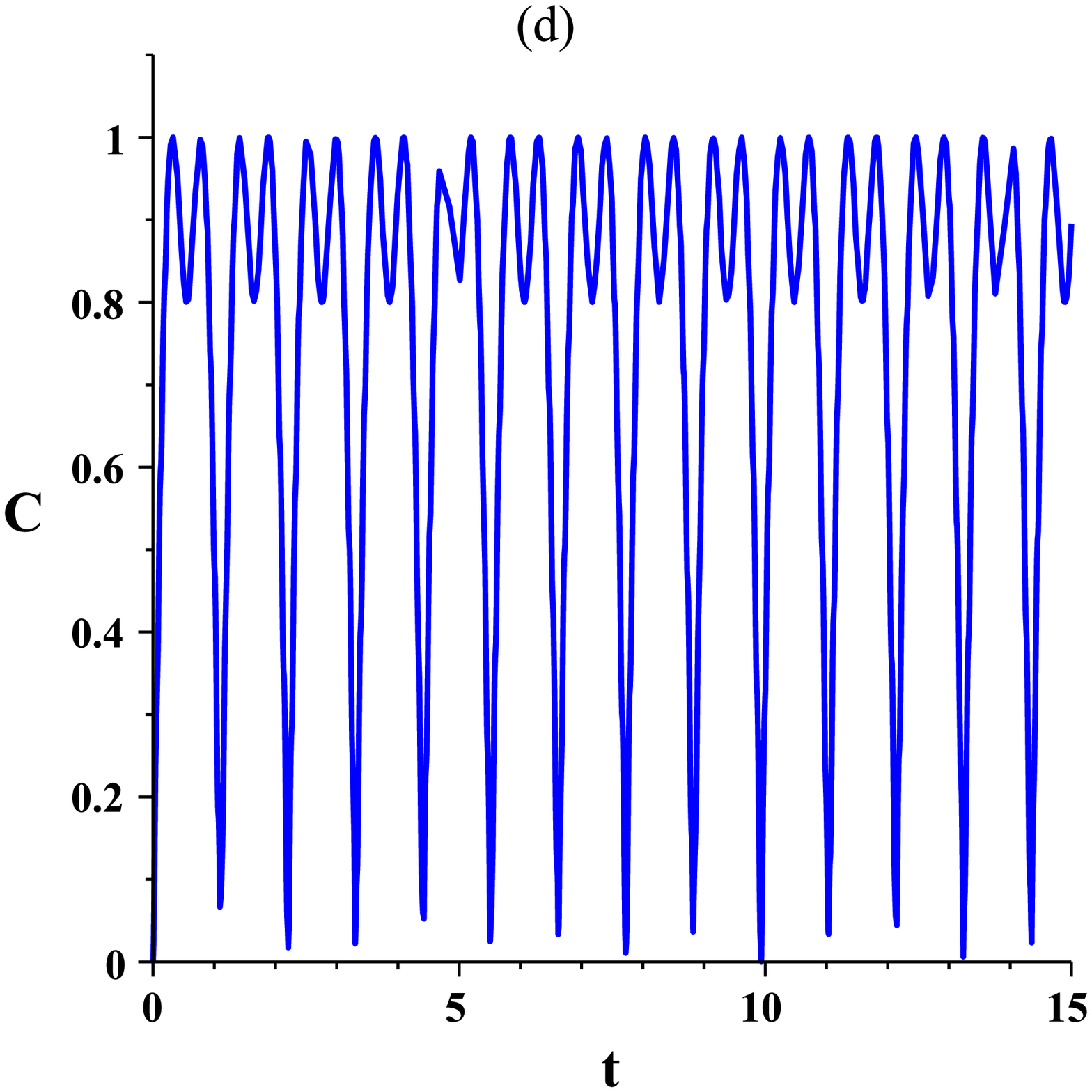}}
   \caption{{\protect\footnotesize First integrability condition: (a) The average magnetic field as a function of time. The time evolution of the concurrence of the state $|++\rangle$ for fixed average magnetic field amplitude $\mu_+=2$, frequency $\beta_+=50$ and initial phase $\phi_+={\pi/50}$ at  (b) $k=0.5 $;  (c) $k=1 $;  (d) $k=2 $.}}
 \label{fig:1}
 \end{minipage}
\end{figure}
we consider the average magnetic field $\omega_{+}(t)$ to be a time-dependent function of the form
\be
\omega_{+}(t)=\mu_{+}\,\sin{\left(\beta_{+}t+\phi_{+}\right)} \;,
\ee
where $\mu_+$, $\beta_+$ and $\phi_+$ are the amplitude, the frequency and the initial phase of $\omega_{+}(t)$, respictively.\\

Starting with the state $|++\rangle$ as an initial state, it evolves in time, when applying the first integrability condition, into
\be
|++\rangle \stackrel{\mathrm{time\, evol.}}\Longrightarrow \left(x_1\, \cos{\th_{10}} - y_1\, \sin{\th_{10}}\right)\, |++\rangle + \left(x_2\, \cos{\th_{10}} - y_2\, \sin{\th_{10}}\right)\, |--\rangle,
\ee
where the coefficients $x's$ and $y's$ can be inferred using
eqs.~(\ref{eig1}), (\ref{intg1}) and (\ref{intg2}) and the the corresponding concurrence is found to be,
\be
C_{++}^{\textsc{ic}} = \left|-2 \sin{2\,\th_{10}}\,\cos{2\,\th_{10}}\, \sin^2{J(t)}
- i\, \sin{2\,\th_{10}}\,\sin{2\,J(t)}\right|\;,
\ee
where,
\be
J(t)=\int_0^{t}\,\eta(t')\,dt'.
\ee
We study the time evolution of the concurrence in fig.~\ref{fig:1} for fixed values of $\mu_+=2$, $\beta_+=50$ and $\phi_+={\pi/50}$  at $k=0.5$, $1$ and $2$. In fig.~\ref{fig:1a} we plot the applied magnetic field $\omega_{+}/\mu_{+}$ against time.
The concurrence exhibits an oscillation behavior which starts from zero magnitude as expected since the initial state is disentangled. In figure~\ref{fig:1b} where $k=0.5$, i.e. $\lambda_{m}(t) <\omega_{+}(t)$, the concurrence oscillation assumes an amplitude $0.8$, but as $k$ increases the value of the amplitude increases too as can be seen in figs.~\ref{fig:1c}. The amplitude reaches the maximum value $1$ at $k=1$ (see fig.~\ref{fig:1c}), but as $k$ increases any further the oscillations becomes distorted as shown in figure~\ref{fig:1} (d), which significantly increases as $\lambda_{m}(t) \gg\omega_{+}(t)$. We didn't find any noticeable change as a result of examining other different initial phases. By comparison with fig.~\ref{fig:1}(a) one can see that the frequency of the concurrence is much smaller than the frequency of the applied field for small values of $k$ but get closer as $k$ increases.
\begin{figure}[htbp]
\begin{minipage}[c]{\textwidth}
 \centering
   \subfigure{\label{fig:2a}\includegraphics[width=5.5cm]{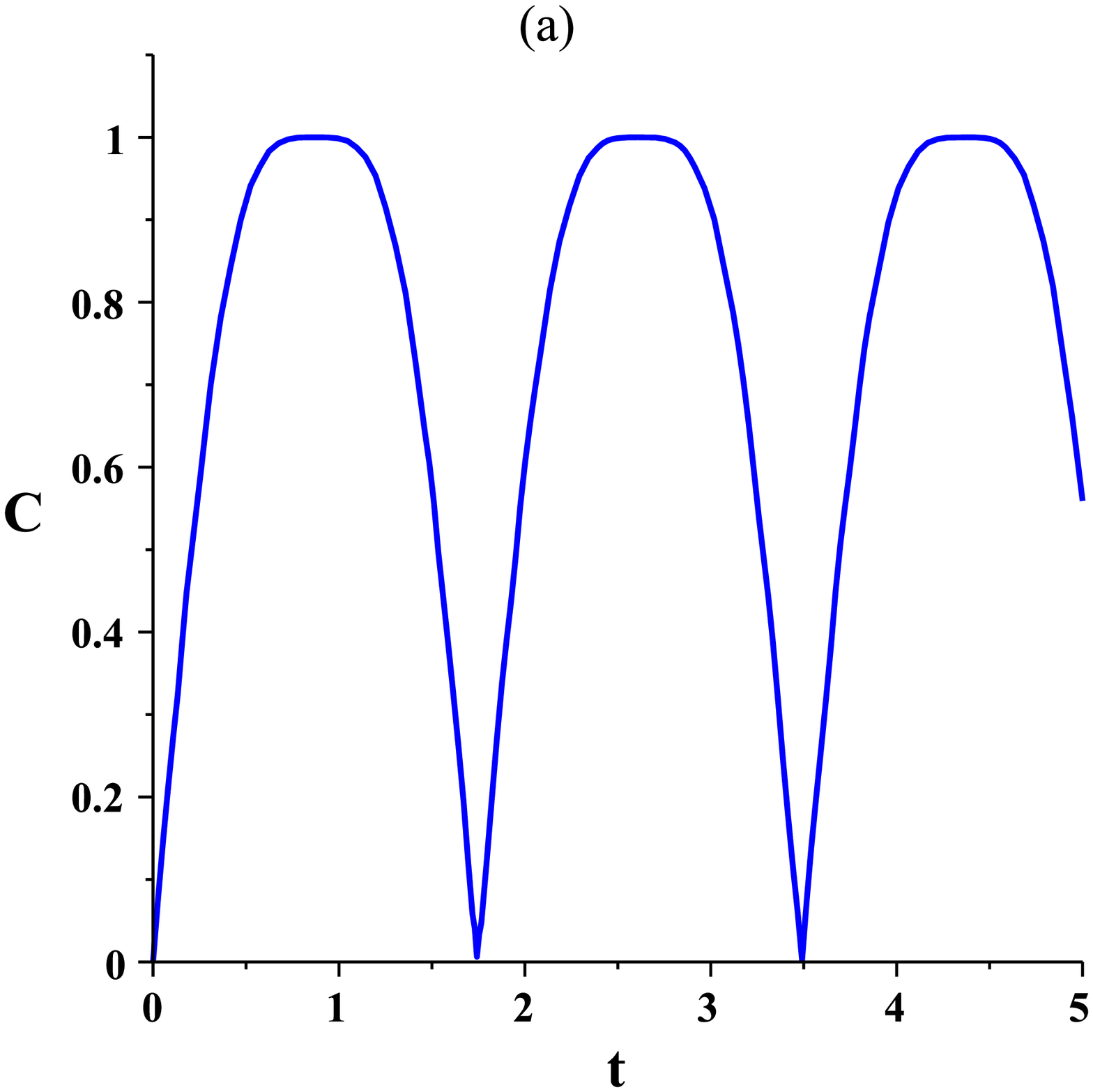}}\quad
   \subfigure{\label{fig:2b}\includegraphics[width=5.5cm]{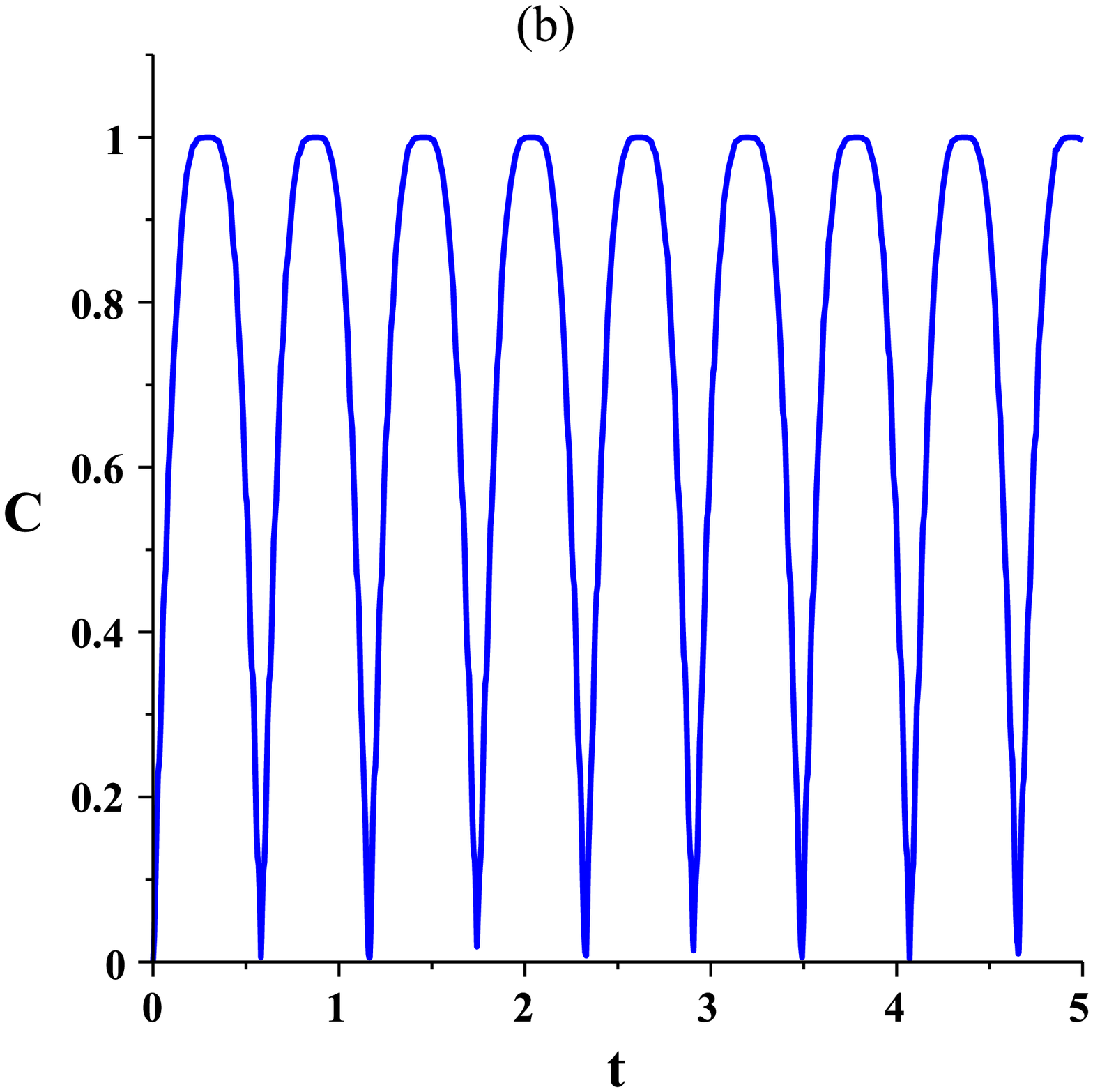}}\\
   \subfigure{\label{fig:2c}\includegraphics[width=5.5cm]{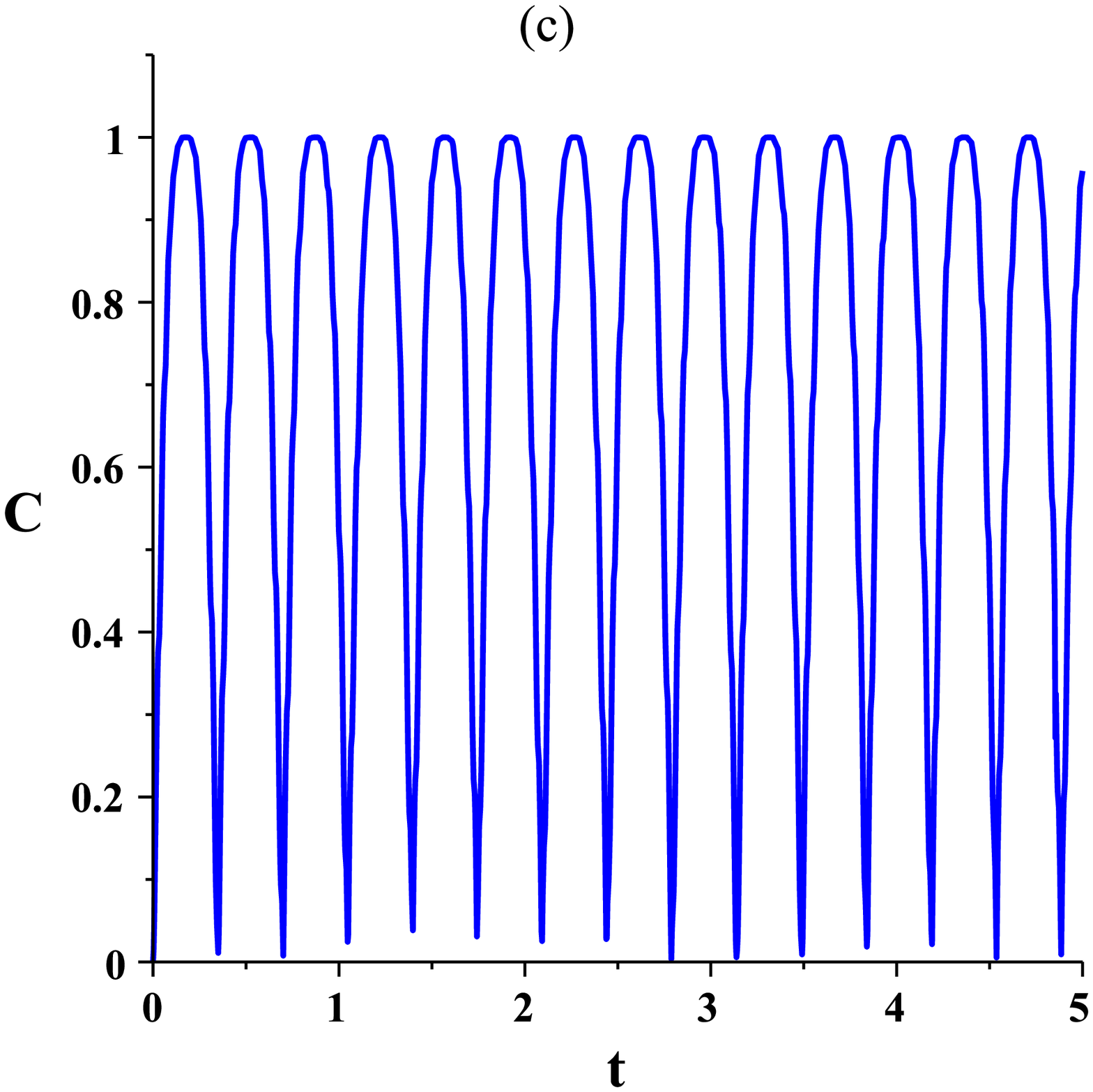}}\quad
   \caption{{\protect\footnotesize First integrability condition: Time evolution of concurrence of the state $|++\rangle$ for fixed average magnetic field frequency $\beta_+=100$, $k=1$ and initial phase $\phi_+={\pi/50}$ at different amplitudes  (a) $\mu_+=2$;  (b) $\mu_+=6$;  (c) $\mu_+=10$.}}
 \label{fig:2}
 \end{minipage}
\end{figure}
In figure~\ref{fig:2}, also we study the concurrence for fixed average magnetic field frequency $\beta_+=100$, $k=1$ and initial phase $\phi_+={\pi/50}$ for different values of $\mu_+=2$, $6$ and $10$. Interestingly, the concurrence exhibits an oscillation behavior where the frequency of the oscillation increases as $\mu_+$ increases.
The time evolution of the concurrence for fixed average magnetic field amplitude $\mu_+=2$, $k=1$ and initial phase $\phi_+={\pi/2}$ at different frequencies has been explored. We didn't notice any change in the concurrence profile as a result of changing the frequency.
\begin{figure}[htbp]
\begin{minipage}[c]{\textwidth}
 \centering
   \subfigure{\label{fig:4a}\includegraphics[width=7.5cm]{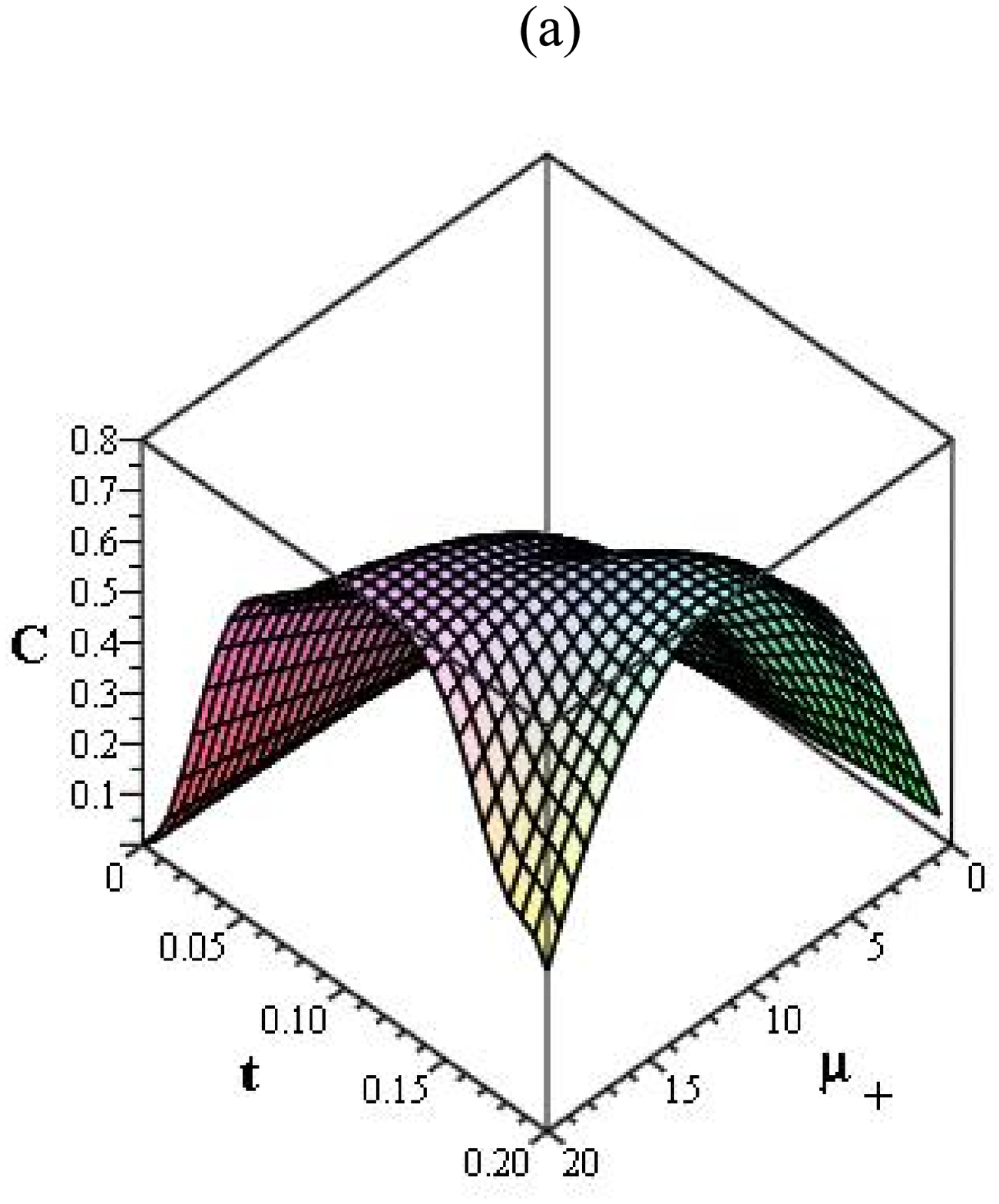}}\quad
   \subfigure{\label{fig:4b}\includegraphics[width=7.5cm]{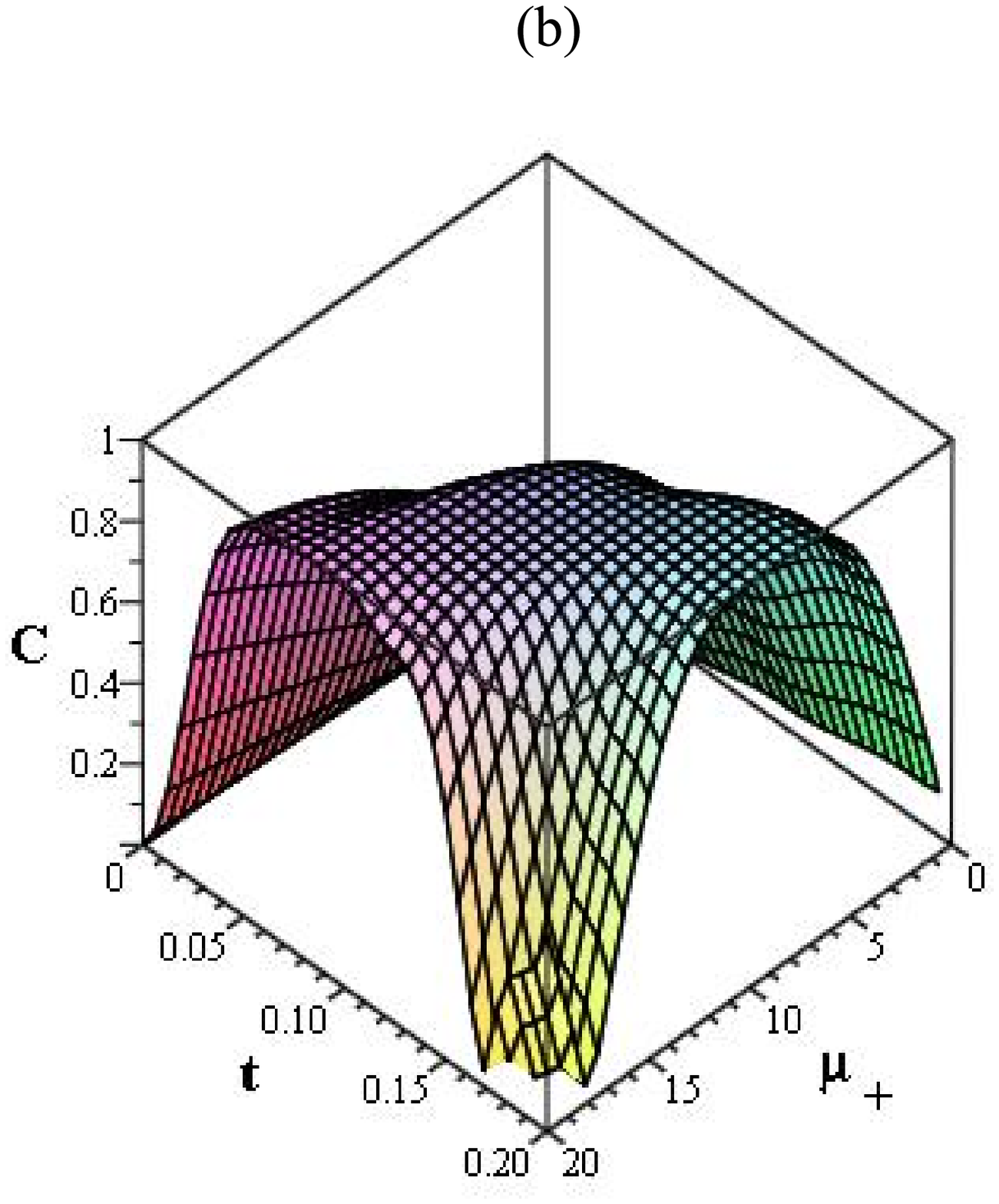}}\\
   \subfigure{\label{fig:4c}\includegraphics[width=7.5cm]{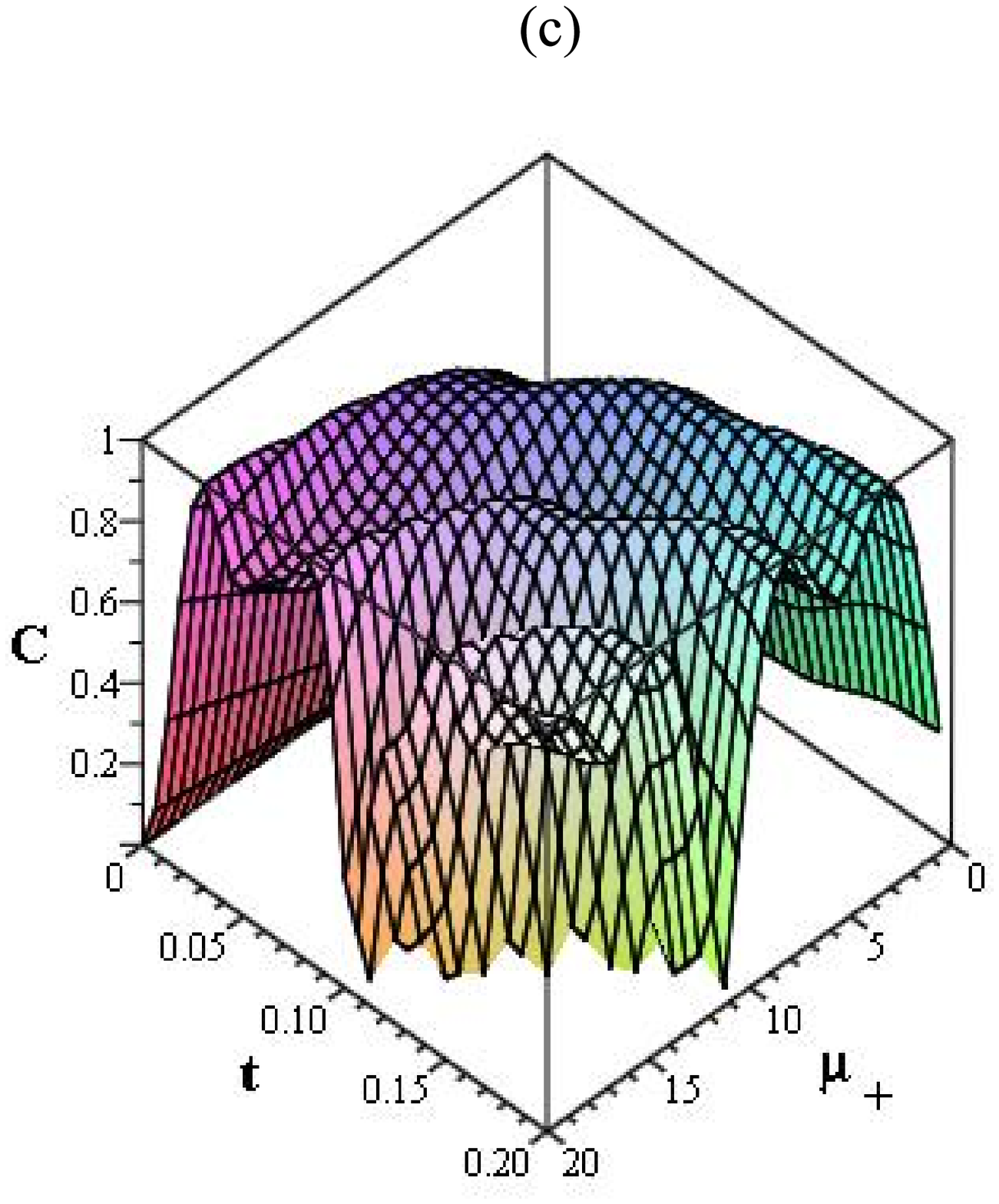}}\quad
   \caption{{\protect\footnotesize First integrability condition: Time evolution of concurrence of the state $|++\rangle$ vs. average magnetic field amplitude $\mu_+$, for frequency $\beta_+=50$ and initial phase $\phi_+={\pi/50}$ at  (a) $k=0.5$;  (b) $k=1$  ;  (c) $k=2$.}}
 \label{fig:4}
 \end{minipage}
\end{figure}
We investigate the time evolution of the concurrence versus average magnetic field amplitude $\mu_+$, for frequency $\beta_+=50$ and initial phase $\phi_+={\pi/50}$ at different values of the parameter $k=0.5$, 1 and 2 in fig.~\ref{fig:4} (a), (b) and (c) respectively. The amplitude of the concurrence oscillation increases as the parameter $k$ increases and its frequency increases also with the average magnetic field amplitude.
Upon starting with entangled state $\frac{1}{\sqrt{2}}\left(|++\rangle + |--\rangle\right)$ as an initial state it evolves in time according to,
\bea
\frac{1}{\sqrt{2}}\left(|++\rangle + |--\rangle\right) &\stackrel{\mathrm{time\, evol.}}\Longrightarrow&
\frac{1}{\sqrt{2}}\,\left[x_1\,\left( \cos{\th_{10}} + \sin{\th_{10}}\right)\, + y_1\,\left( \cos{\th_{10}} - \sin{\th_{10}}\right)\right] |++\rangle \nn\\
&&
+ \frac{1}{\sqrt{2}}\,\left[x_2\,\left( \cos{\th_{10}} + \sin{\th_{10}}\right)\, + y_2\,\left( \cos{\th_{10}} - \sin{\th_{10}}\right)\right] |--\rangle,\nn\\
\eea
while the corresponding entanglement is,
\be
C_{++--\mbox{\textsc{s}}}^{\textsc{ic}} = \left|- i\, \sin{2\,\th_{10}}\, \sin{2\,J(t)}
+ \sin^2{2\,\th_{10}}\,\cos{2\,J(t)}
+ \cos^2{2\,\th_{10}}\right|.
\ee
\begin{figure}[htbp]
\begin{minipage}[c]{\textwidth}
 \centering
   \subfigure{\label{fig:5a}\includegraphics[width=5.5cm]{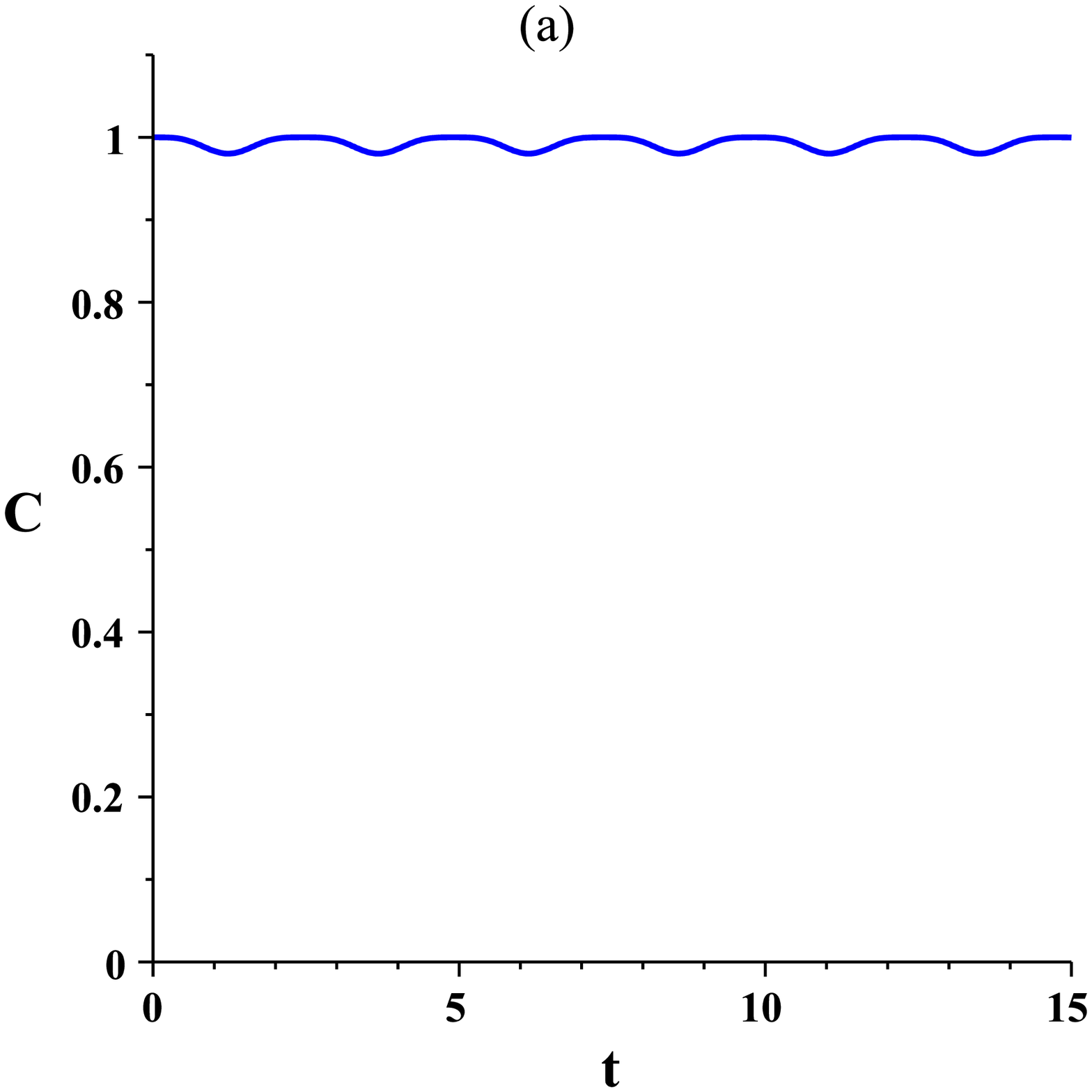}}\quad
   \subfigure{\label{fig:5b}\includegraphics[width=5.5cm]{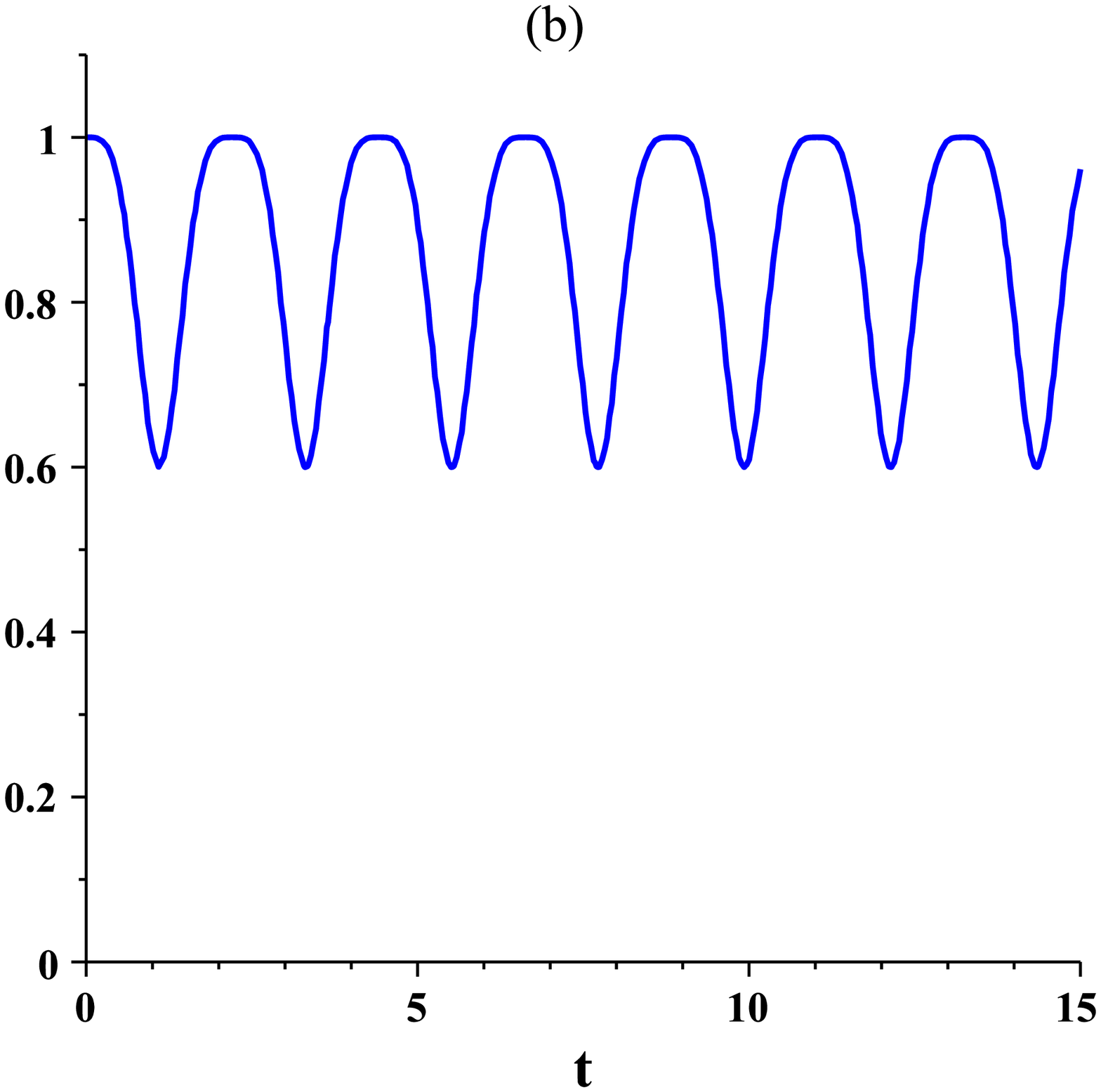}}\\
   \subfigure{\label{fig:5c}\includegraphics[width=5.5cm]{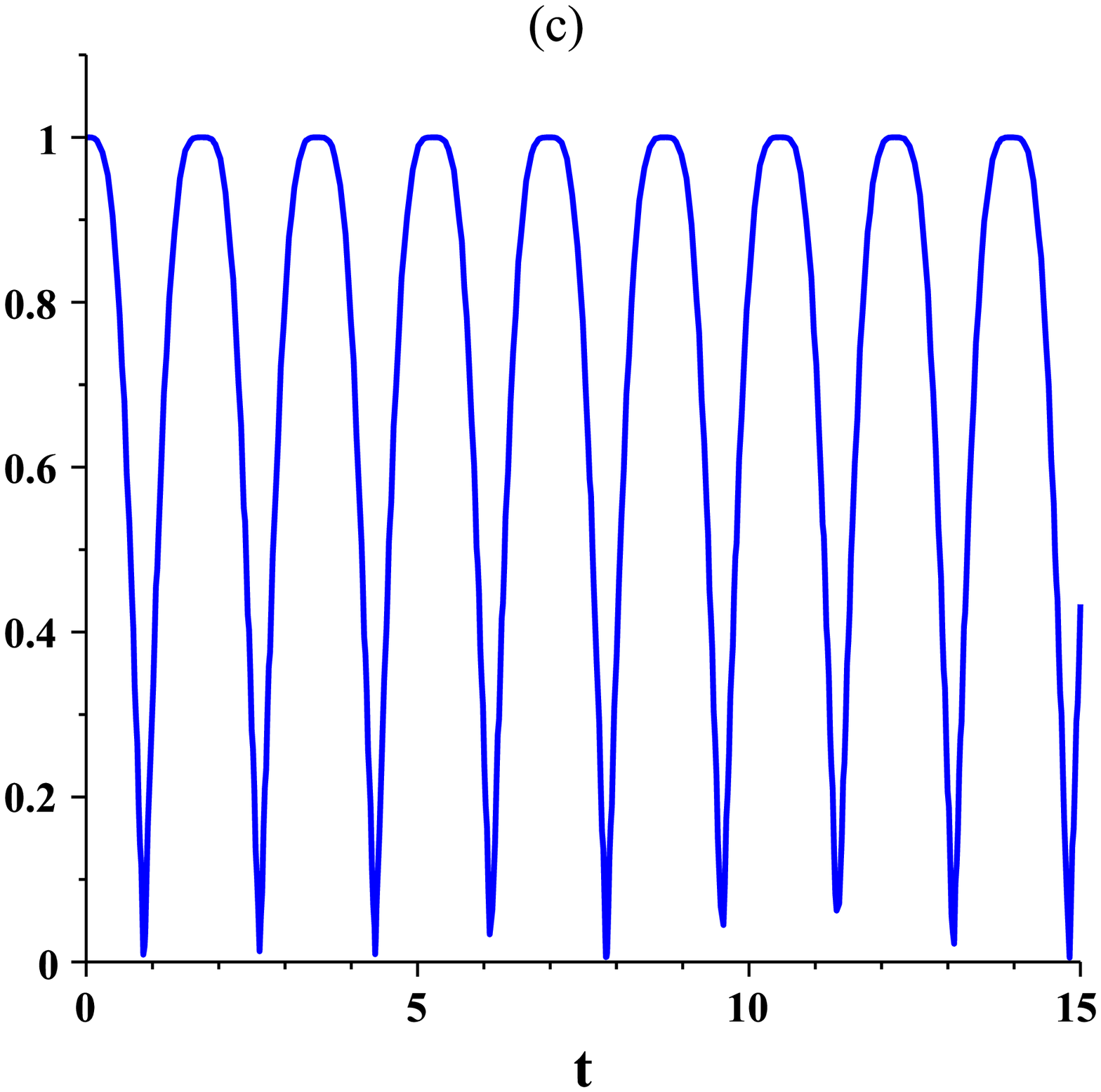}}\quad
   \subfigure{\label{fig:5d}\includegraphics[width=5.5cm]{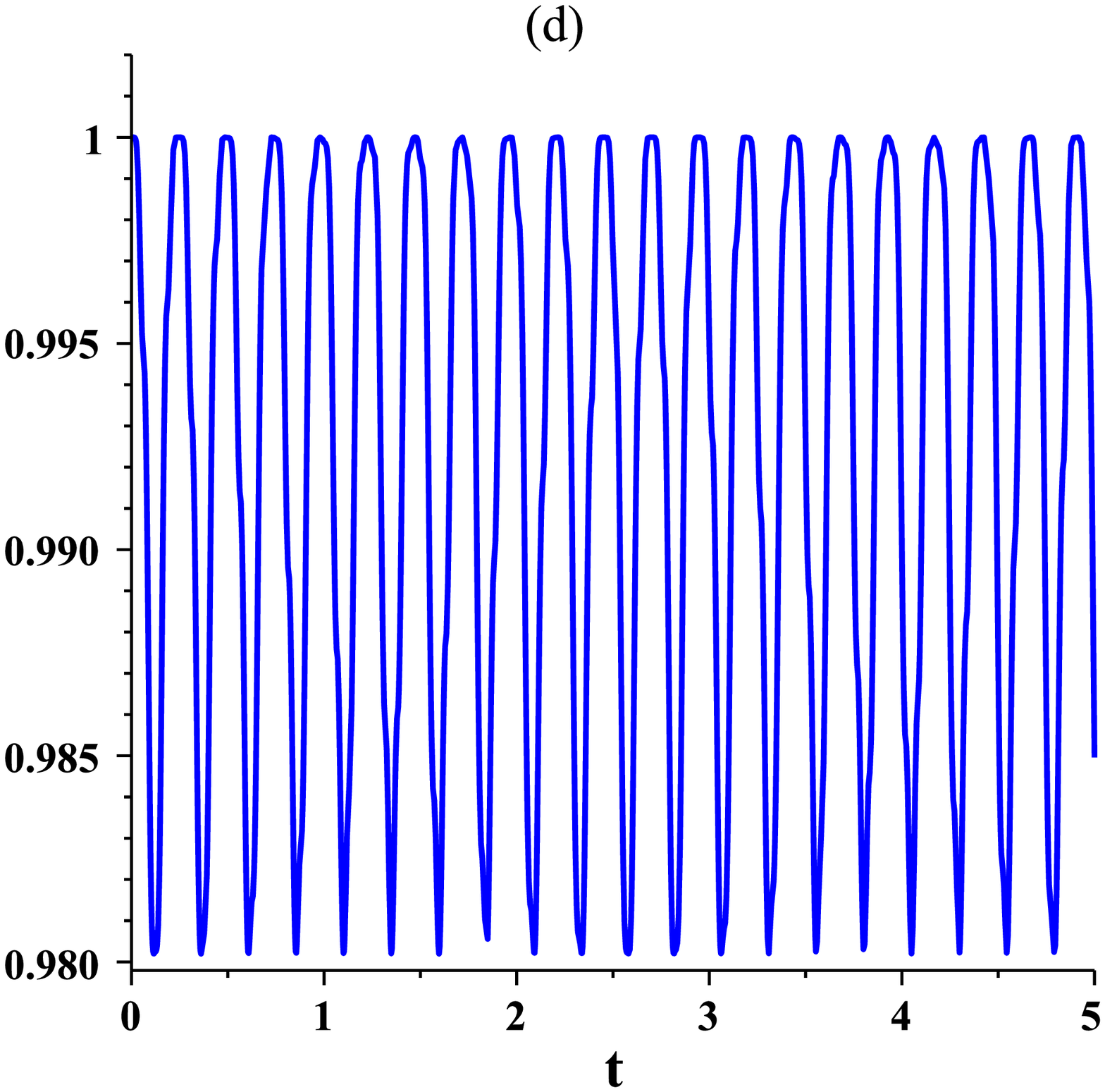}}
   \caption{{\protect\footnotesize First integrability condition: Time evolution of concurrence of the state $\frac{1}{\sqrt{2}}\left(|++\rangle + |--\rangle\right)$ for fixed average magnetic field amplitude $\mu_+=2$, frequency $\beta_+=50$ and initial phase $\phi_+={\pi/50}$ at    (a) $k=0.1 $;  (b) $k=0.5 $;  (c) $k=1 $;  (d) $k=10 $.}}
 \label{fig:5}
 \end{minipage}
\end{figure}
In fig.~\ref{fig:5}, we study the time evolution of the concurrence for fixed average magnetic field amplitude $\mu_+=2$, frequency  $\beta_+=50$ and initial phase $\phi_+={\pi/50}$ at $k=0.1$, $0.5$, $1$ and $10$. As can be noticed, the concurrence exhibits an oscillatory behavior starting with an initial value of 1 as expected for a maximally entangled initial state. At $k=0.1$ $(\lambda_{m}(t) <\omega_{+}(t))$, shown in fig~\ref{fig:5a}, the concurrence amplitude is very small but as $k$ increases the amplitude increases too reaching its maximum value of 1 at $k=1$, as shown in figs.~\ref{fig:5c}. The amplitude gets damped again and the frequency increases as $k$ increases any further as can be concluded from fig.\ref{fig:5d}. The same state was considered at different initial phases and frequencies with no significant change in the concurrence behavior.
The effect of the average magnetic field amplitude has been examined. The concurrence exhibits an oscillatory behavior with a frequency that increases as $\mu_+$ increases similar to the disentangled case.

 The concurrences corresponding to a system which is initially in in the state  $|--\rangle$ or $\frac{1}{\sqrt{2}}\left(|++\rangle - |--\rangle\right)$ can be straightforwardly calculated leading respectively to
\be
C_{--}^{\textsc{ic}} = \left|2 \sin{2\,\th_{10}}\,\cos{2\,\th_{10}}\, \sin^2{J(t)}
- i\, \sin{2\,\th_{10}}\,\sin{2\,J(t)}\right|,
\ee
and
\be
C_{++--\mbox{\textsc{a}}}^{\textsc{ic}} = \left|- i\, \sin{2\,\th_{10}}\, \sin{2\,J(t)}
-\sin^2{2\,\th_{10}}\,\cos{2\,J(t)}
- \cos^2{2\,\th_{10}}\right| \;.
\ee

\subsection{Entanglement formula for the second integrability condition}
The formulae for time evolution using the second integrability condition are little bit complicated see eqs.~(\ref{eqic23}), (\ref{eqic24}) and (\ref{eqic25}). However, the expressions can be written in a simple manner if they are broken up into sub-expressions that have simpler structures. The relevant sub-expressions are,
\bea
&&\mbox{Im}\left(x_1\,x_2\right)= \mbox{Im}\left(y_1\,y_2\right)= {\kappa\,\sin^2{\d_1}\,\cos{2\,\th_1} \over 1 + \kappa^2} -{1\over 2} {\mbox{sgn}(k)\,\sin{2\,\th_1}\,\sin{2\d_1}\over \sqrt{1+\kappa^2}},\nn\\
&& \mbox{Re}\left(x_1\,x_2\right)= -\mbox{Re}\left(y_1\,y_2\right)= {1\over 2}\,\sin{2\,\th_1}\,\cos{2\d_1} -{1\over 2}\,{|\kappa|\,\sin{2\,\d_1}\,\cos{2\,\th_1} \over \sqrt{1 + \kappa^2}},\nn\\
&& x_1\,y_2 + y_1\,x_2 = {|\kappa|\,\sin{2\,\th_1}\,\sin{2\,\d_1} \over \sqrt{1 + \kappa^2}} + \cos^2{\d_1}\,\cos{2\,\th_1} + \sin^2{\d_1}\,\cos{2\,\th_1}\,{1-\kappa^2 \over 1 + \kappa^2},
\label{partxy}
\eea
where $\mbox{Re}$ and \mbox{Im} denote respectively the real and imaginary part.
\begin{figure}[htbp]
\begin{minipage}[c]{\textwidth}
 \centering
   \subfigure{\label{fig:8a}\includegraphics[width=5.5cm]{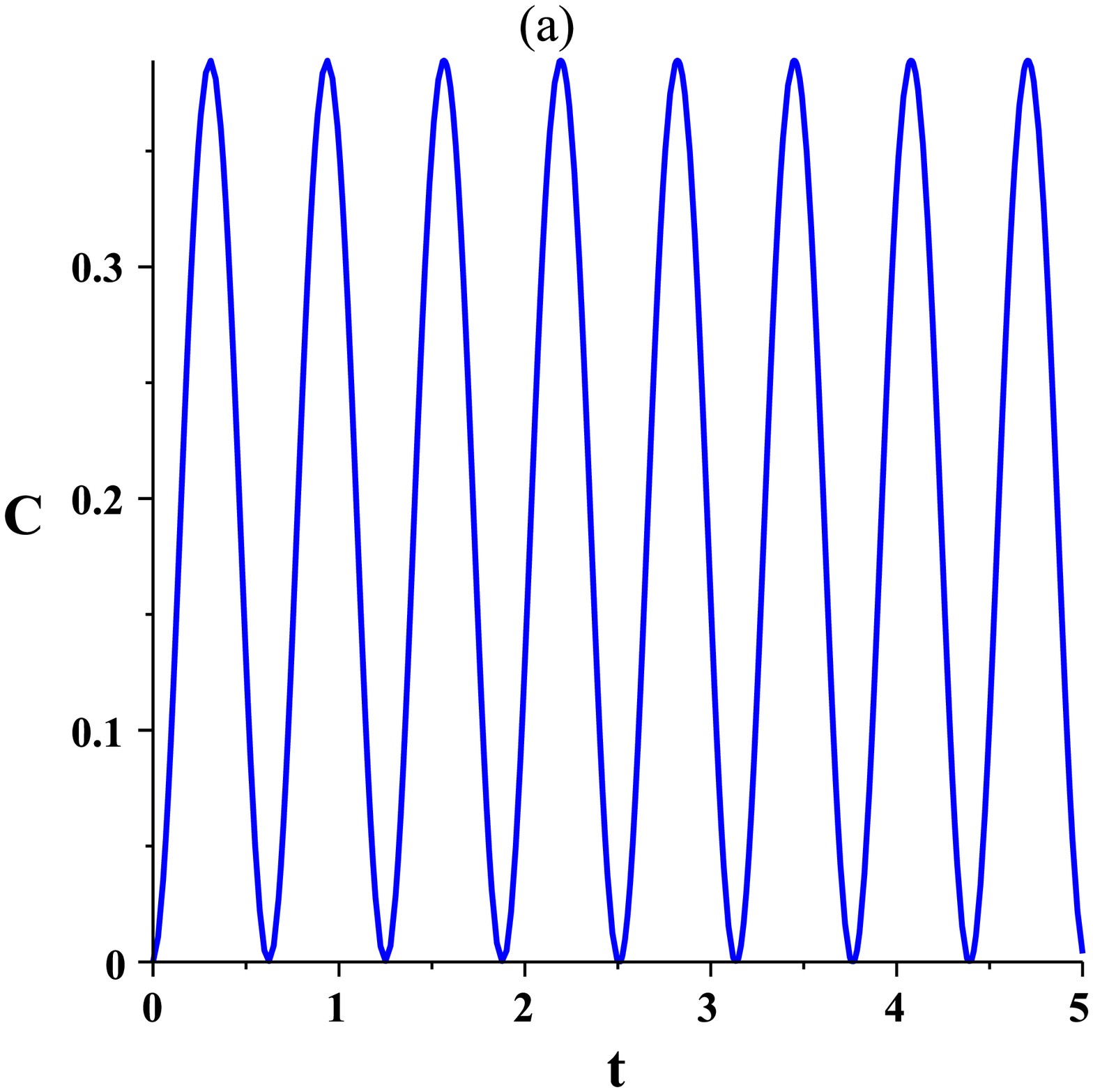}}\quad
   \subfigure{\label{fig:8b}\includegraphics[width=5.5cm]{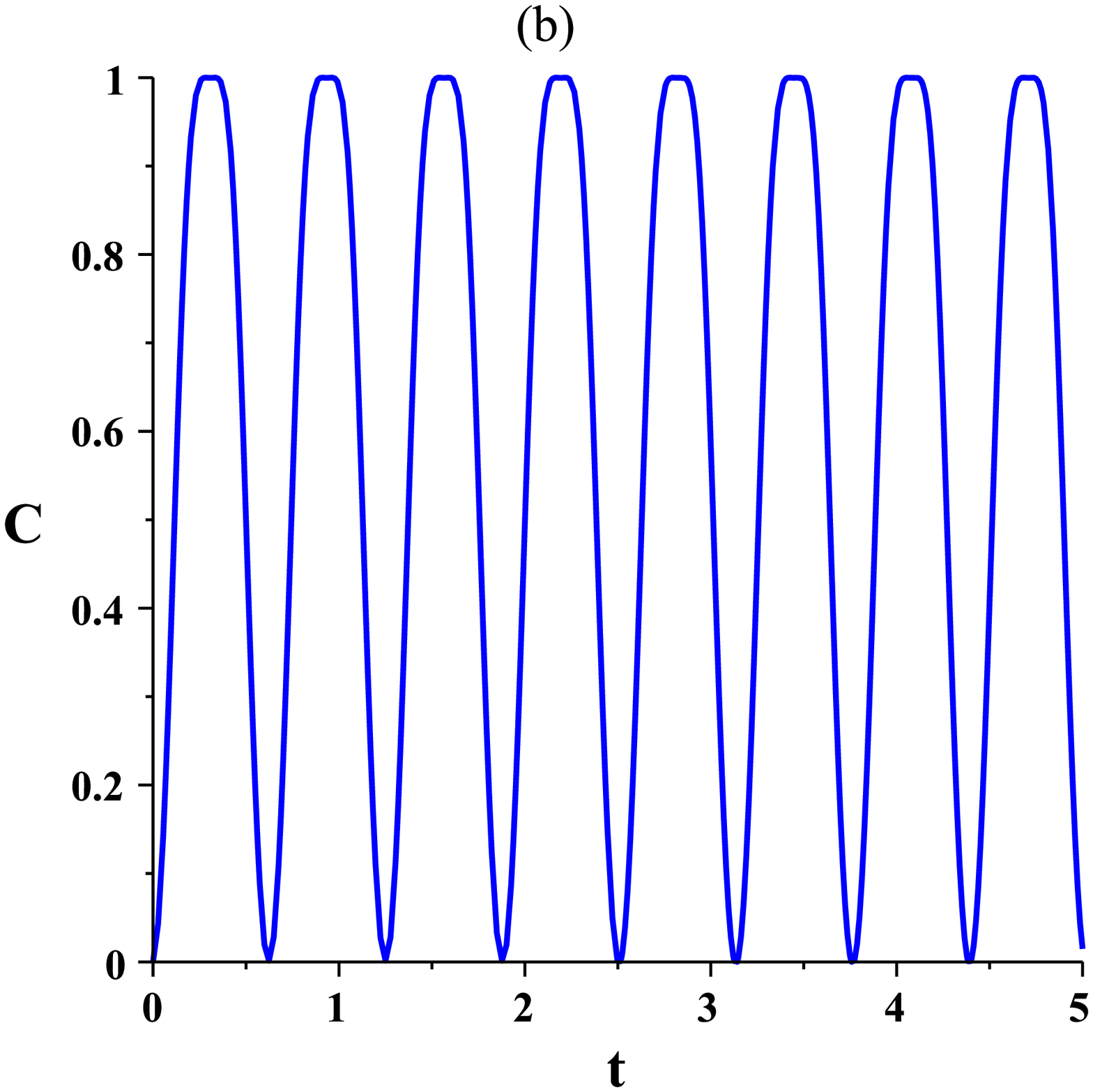}}\\
   \subfigure{\label{fig:8c}\includegraphics[width=5.5cm]{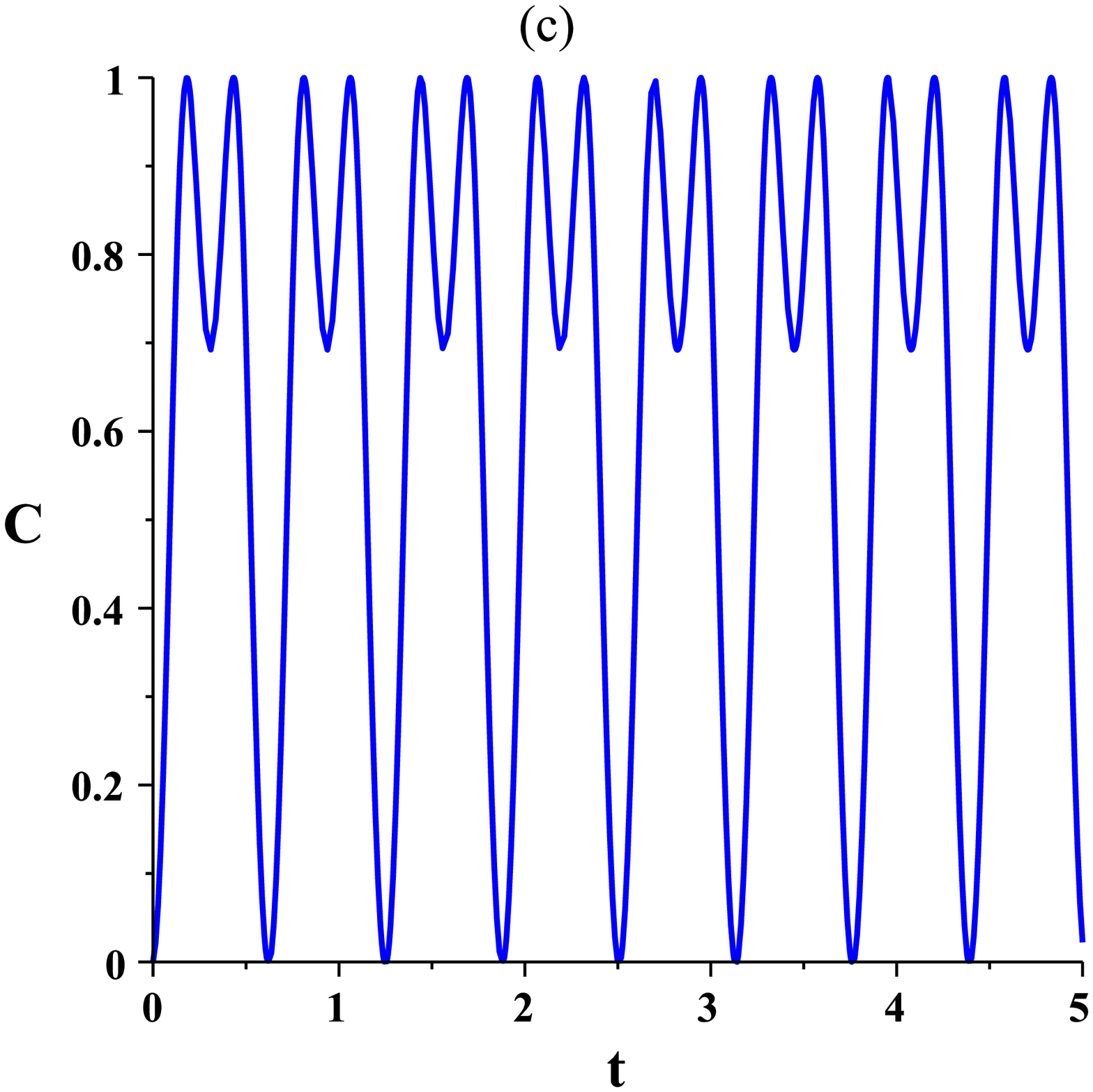}}
   \caption{{\protect\footnotesize Second integrability condition: Time evolution of concurrence of the state $|++\rangle$ for fixed Heisenberg interaction $(\la_m(t))$ frequency $\beta_{m}=10$, $\kappa=0.1$, $\theta_{10}=\pi/4$ and initial phase $\phi_{m}={\pi/50}$ at different amplitudes   (a) $\mu_{m}=1 $;  (b)  $\mu_{m}=4 $;  (c)  $\mu_{m}=6 $.}}
 \label{fig:8}
 \end{minipage}
\end{figure}

With the aid of eqs.~(\ref{conpsiI}) and (\ref{partxy}), the entanglement associated with time evolved state starting from the disentangled states $|++\rangle$  and $|--\rangle$ amount respectively to be
\bea
C_{++}^{\textsc{iic}} &=& 2\,\left|x_1\,x_2\,\cos^2{\th_{10}} + y_1\,y_2\,\sin^2{\th_{10}}
- \sin{\th_{10}}\,\cos{\th_{10}}\,\left(x_1\,y_2 + y_1\,x_2\right)\right|,\nn\\
 &=& \left| 2 \cos{2\,\th_{10}}\,\mbox{Re}\left(x_1\,x_2\right) + 2\,I\,\mbox{Im}\left(x_1\,x_2\right)
- \sin{2\,\th_{10}}\,\left(x_1\,y_2 + y_1\,x_2\right)\right|,\nn\\\nn\\
C_{--}^{\textsc{iic}} &=& \left| -2 \cos{2\,\th_{10}}\,\mbox{Re}\left(x_1\,x_2\right) + 2\,I\,\mbox{Im}\left(x_1\,x_2\right)
+ \sin{2\,\th_{10}}\,\left(x_1\,y_2 + y_1\,x_2\right)\right|\;.
\label{cdic2}
\eea
\begin{figure}[htbp]
\begin{minipage}[c]{\textwidth}
 \centering
   \subfigure{\label{fig:9a}\includegraphics[width=5.5cm]{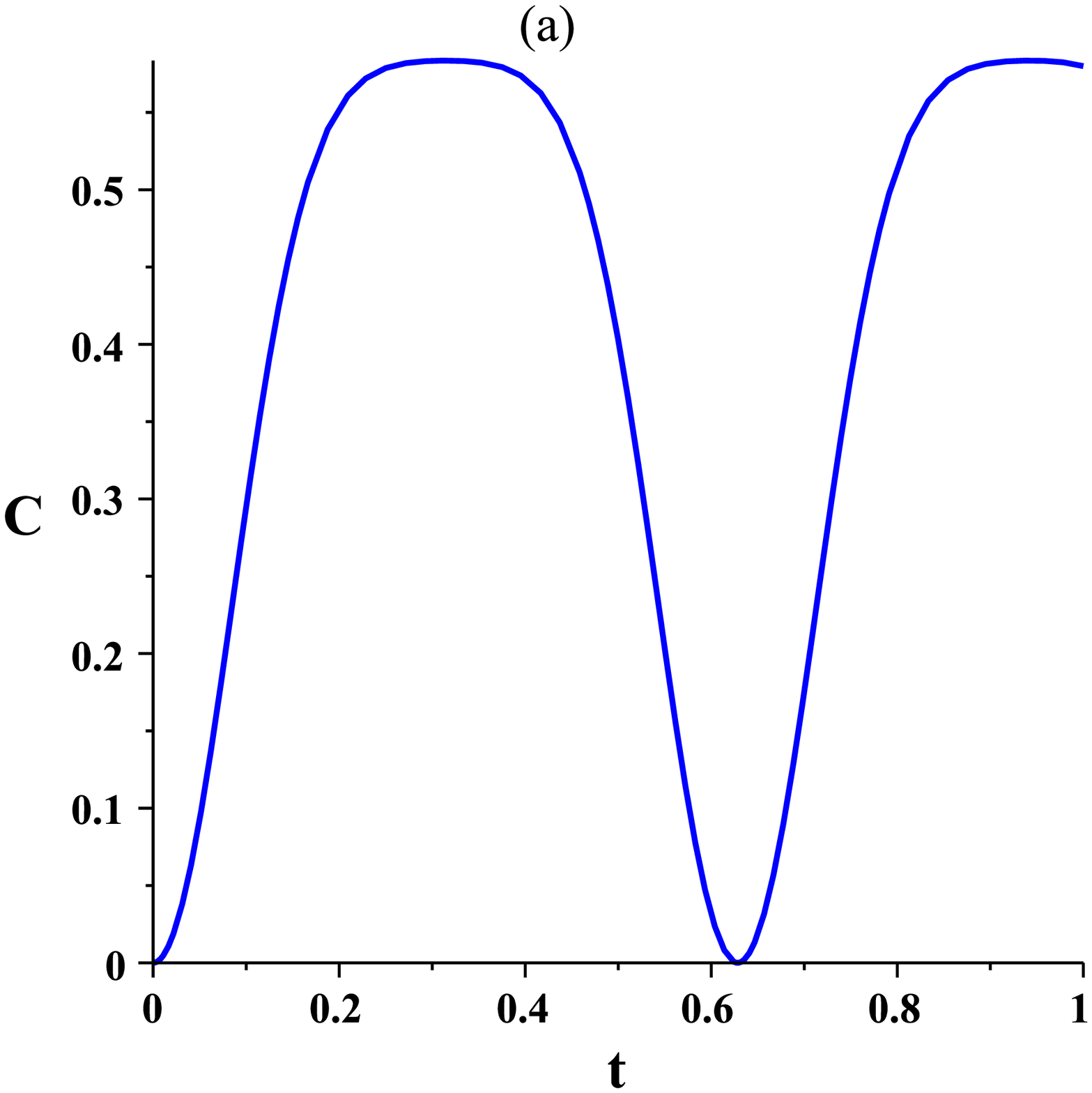}}\quad
   \subfigure{\label{fig:9b}\includegraphics[width=5.5cm]{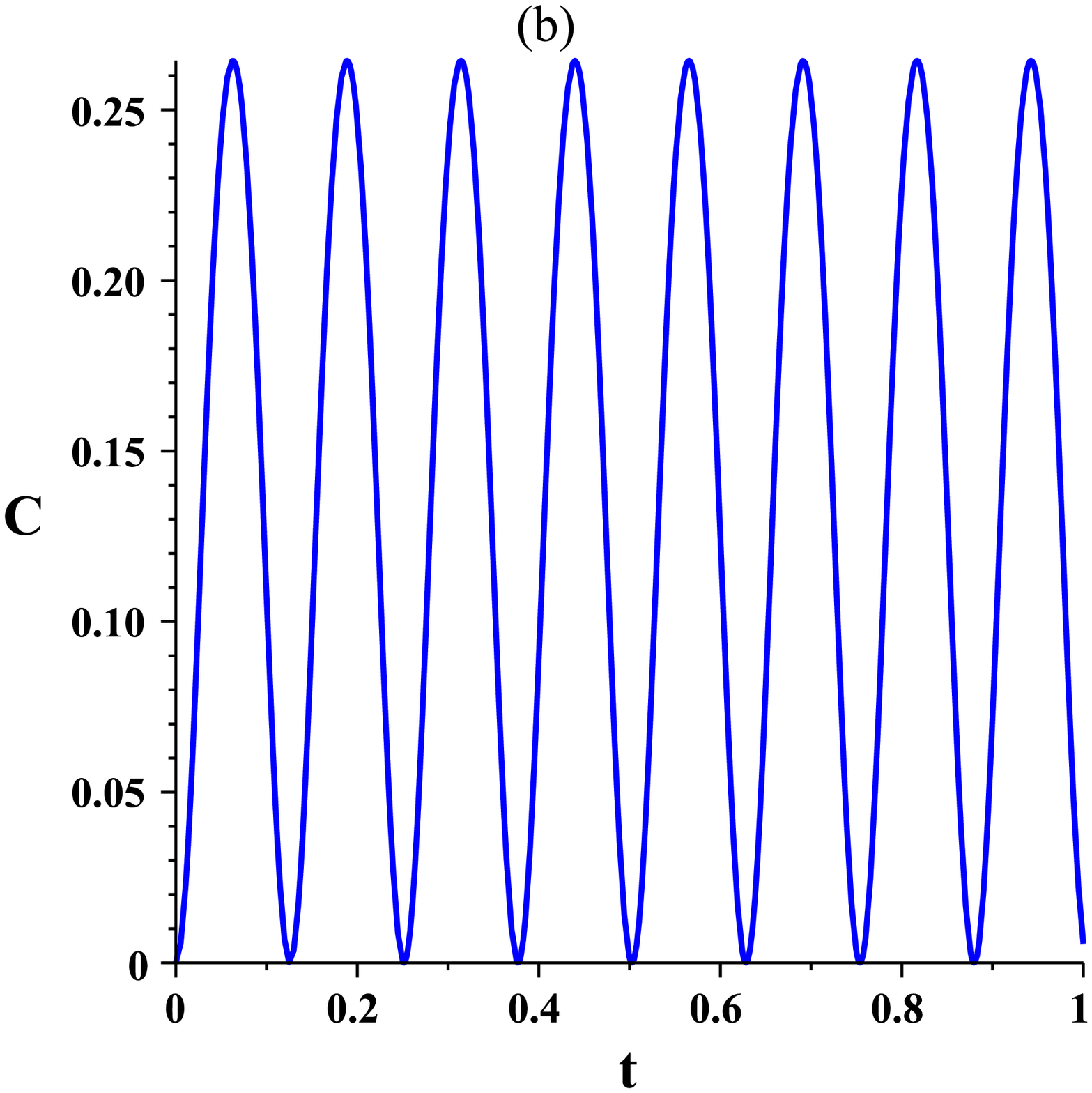}}\\
   \subfigure{\label{fig:9c}\includegraphics[width=5.5cm]{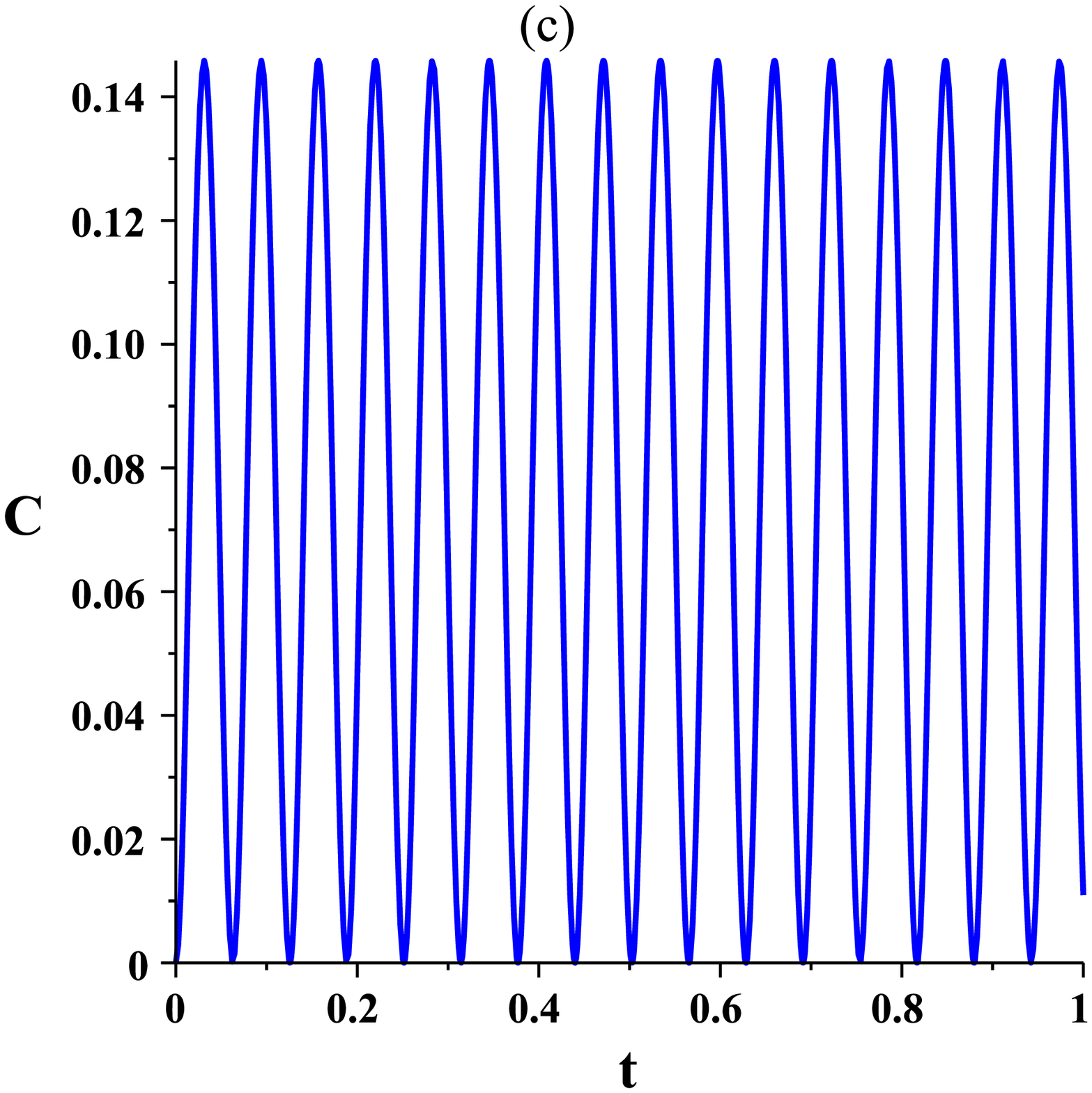}}
   \caption{{\protect\footnotesize Second integrability condition: Time evolution of concurrence of the state $|++\rangle$ for fixed amplitude of Heisenberg interaction, $(\la_m(t))$, $\mu_{m}=4 $, $\kappa=0.1$, $\theta_{10}=0$ and initial phase $\phi_{m}=0$ at different frequencies   (a) $\beta_{m}=10$;  (b)$\beta_{m}=50$ ;  (c)  $\beta_{m}=100$.}}
 \label{fig:9}
 \end{minipage}
\end{figure}


The resulting entanglement associated with the initially  maximally entangled states
 $\frac{1}{\sqrt{2}}\left(|++\rangle + |--\rangle\right)$ and $\frac{1}{\sqrt{2}}\left(|++\rangle - |--\rangle\right)$ as functions of time are given by,
\bea
C_{++--\mbox{\textsc{s}}}^{\textsc{iic}}&=& \left|x_1\,x_2\,\left(1+\sin{2\,\th_{10}}\right) + y_1\,y_2\,\left(1-\sin{2\,\th_{10}}\right)
+ \cos{2\,\th_{10}}\,\left(x_1\,y_2 + y_1\,x_2\right)\right|,\nn\\
&=& \left| 2 \sin{2\,\th_{10}}\,\mbox{Re}\left(x_1\,x_2\right) + 2\,I\,\mbox{Im}\left(x_1\,x_2\right)
+ \cos{2\,\th_{10}}\,\left(x_1\,y_2 + y_1\,x_2\right)\right|,\nn \\\nn\\
C_{++--\mbox{\textsc{a}}}^{\textsc{iic}}&=&
\left| -2 \sin{2\,\th_{10}}\,\mbox{Re}\left(x_1\,x_2\right) + 2\,I\,\mbox{Im}\left(x_1\,x_2\right)
- \cos{2\,\th_{10}}\,\left(x_1\,y_2 + y_1\,x_2\right)\right|
\label{cmic2}
\eea


Note that the entanglement for state evolving initially from $|--\rangle$  can be obtained from that of  $|++\rangle$ by using the symmetry prescribed in eqs.~(\ref{sym3}, \ref{flip2}). this means that,
\bea
&&\th_{10} \rightarrow {\pi\over 2} -\th_{10},\;\;  \th \rightarrow {\pi\over 2} -\th,\nn\\
&&x_1 \leftrightarrow x_2,\;\;   y_1 \leftrightarrow - y_2,\;\;  k\rightarrow -k \;,
\eea
While the results in eq.(\ref{cmic2}), reveals that the entanglement of state  evolving initially from $\frac{1}{\sqrt{2}}\left(|++\rangle - |--\rangle\right)$, can be obtained from that of $\frac{1}{\sqrt{2}}\left(|++\rangle + |--\rangle\right)$, by the following replacement
\bea
\cos{\th_{10}}\Rightarrow - \sin{\th_{10}}, && \sin{\th_{10}}\Rightarrow \cos{\th_{10}},
\label{orthoconv}
\eea
which is the appropriate replacement to transform a state to its orthogonal counterpart in the same subspace. The same rule in eq.~(\ref{orthoconv}) can be applied to the states
$|++\rangle$ and $|--\rangle$.

Again, suppose we consider $|++\rangle$ as an initial state, we investigate the time evolution of the concurrence in fig.~\ref{fig:8} for a fixed Heisenberg interaction frequency $\beta_{m}=10$, $\kappa=0.1$, $\theta_{10}=\pi/4$ and initial phase $\phi_{m}={\pi/50}$ at different amplitudes $\mu_{m}=1$, $4$ and $6$. The concurrence shows an oscillatory behavior, where as $\mu_{m}$ increases the oscillation amplitude increases, but becomes distorted for large values of $\mu_{m}$. The effect of the frequency $\beta_{m}$ on the concurrence has been examined. Increasing $\beta_{m}$ causes the concurrence frequency to increase whereas the amplitude decreases. For fixed parameters, there is no effect of $\kappa$ on the concurrence.

The impact of the angle $\theta_{10}=0$ and initial phase $\phi_{m}=0$ has been investigated for a fixed Heisenberg interaction frequency $\beta_{m}=10$, $\kappa=0.1$ at different amplitudes $\mu_{m}=1$, $4$ and $6$. We got the same results as in the case of $\theta_{10}=\pi/4$ and initial phase $\phi_{m}={\pi/50}$.

The time evolution of the concurrence is depicted for fixed parameter values $\mu_{m}=4$, $\kappa=0.1$, $\theta_{10}=0$ and initial phase $\phi_{m}=0$ at different frequencies $\beta_{m}=10$, $50$ and $100$ in figs.~\ref{fig:9a}, (b) and (c), respectively. The concurrence shows an oscillatory behavior with a frequency that increases and a decresing amplitude as $\beta_{m}$ increases.
For fixed parameters, we have tested the effect of $\kappa$ which when increases the amplitude of the concurrence increases.

\begin{figure}[htbp]
\begin{minipage}[c]{\textwidth}
 \centering
   \subfigure{\label{fig:10a}\includegraphics[width=7.5cm]{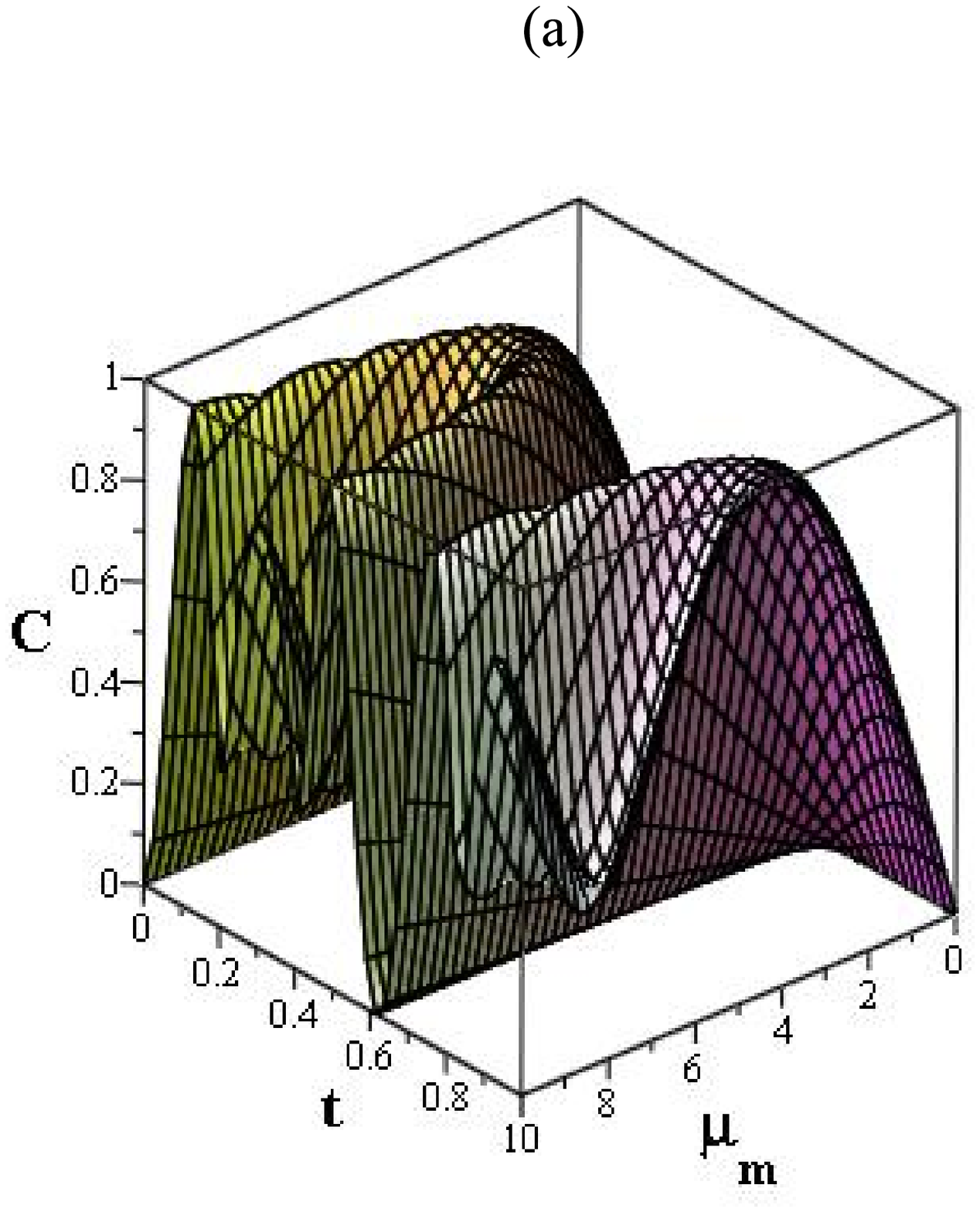}}\quad
   \subfigure{\label{fig:10b}\includegraphics[width=7.5cm]{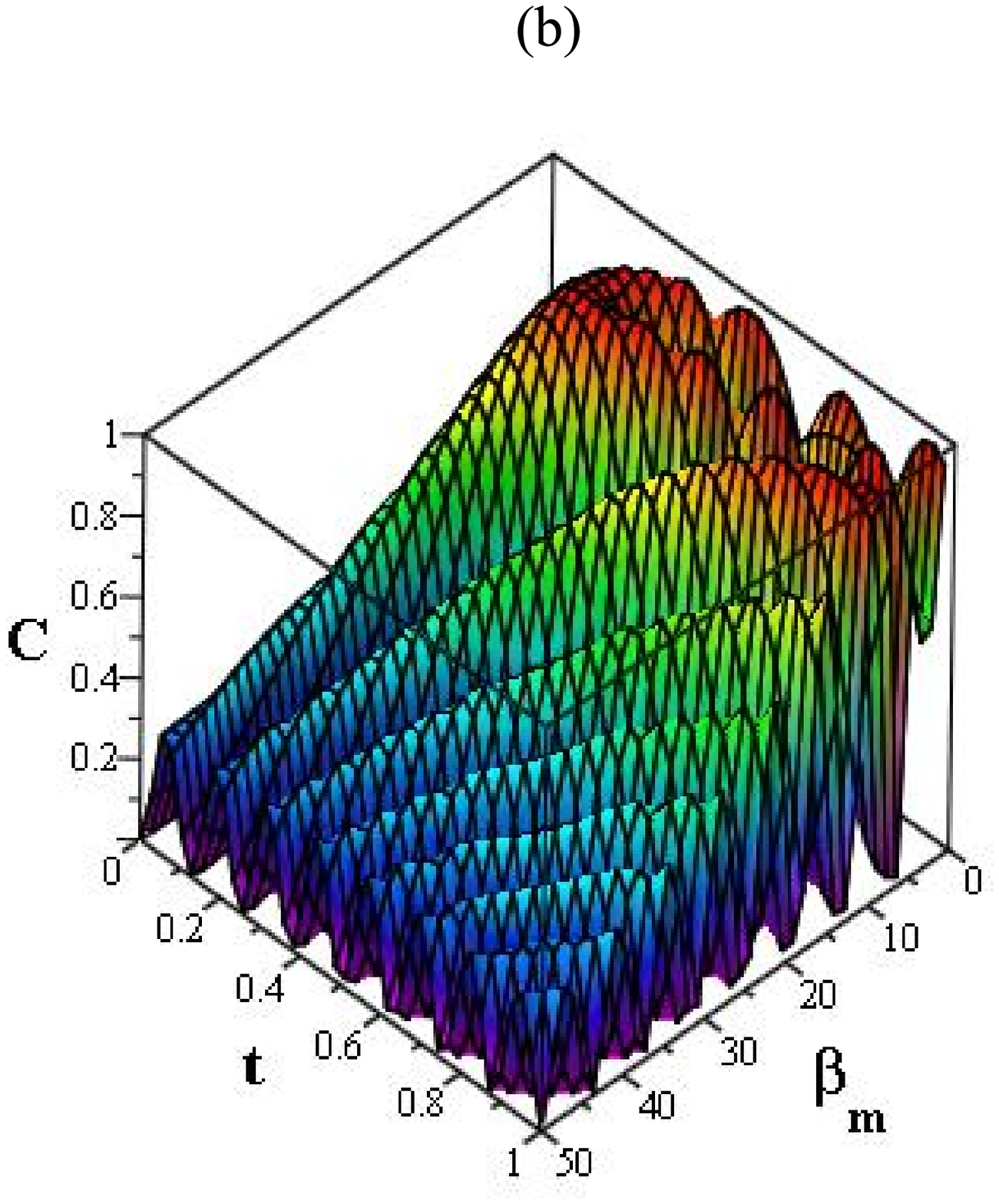}}
   \caption{{\protect\footnotesize Second integrability condition:  Time evolution of concurrence of the state $|++\rangle$ (a) as a function of  $\mu_{m}$, for fixed frequency $\beta_{m}=10$, $\kappa=0.1$, $\theta_{10}=\pi/4$ and initial phase $\phi_{m}={\pi/50}$. (b) as a function of $\beta_{m}$, for fixed amplitude $\mu_{m}=4$, $\kappa=0.1$, $\theta_{10}=\pi/4$ and initial phase $\phi_{m}={\pi/50}$. Notice that $\la_m(t)= \mu_m \sin{\left(\beta_m t + \phi_m\right)}$. }}
 \label{fig:10}
 \end{minipage}
\end{figure}

The time evolution of concurrence versus the Heisenberg interaction amplitude $\mu_{m}$ is depicted in fig.~\ref{fig:10a}, for fixed frequency $\beta_{m}=10$, $\kappa=0.1$, $\theta_{10}=\pi/4$ and initial phase $\phi_{m}={\pi/50}$. The concurrence exhibits an oscillatory behavior, which gets distorted as the magnitude $\mu_{m}$ increases. In fig.~\ref{fig:10b} The time evolution of concurrence versus the Heisenberg interaction frequency $\beta_{m}$ is investigated for fixed amplitude $\mu_{m}=4$, $\kappa=0.1$, $\theta_{10}=\pi/4$ and initial phase $\phi_{m}={\pi/50}$. Clearly the concurrence oscillation frequency increases and its amplitude decreases as $\beta_{m}$ increases.

\begin{figure}[htbp]
\begin{minipage}[c]{\textwidth}
 \centering
   \subfigure{\label{fig:11a}\includegraphics[width=5.5cm]{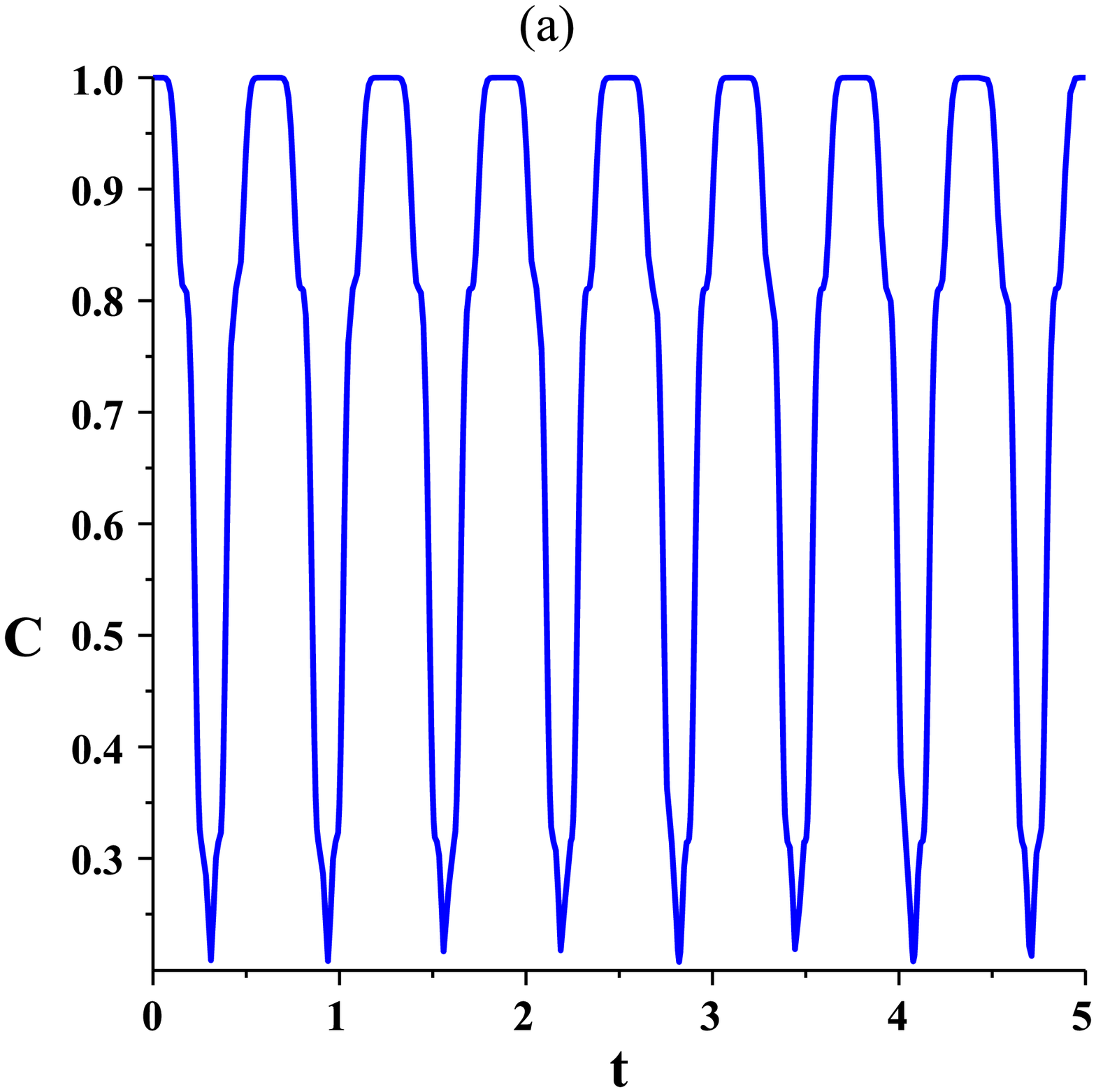}}\quad
   \subfigure{\label{fig:11b}\includegraphics[width=5.5cm]{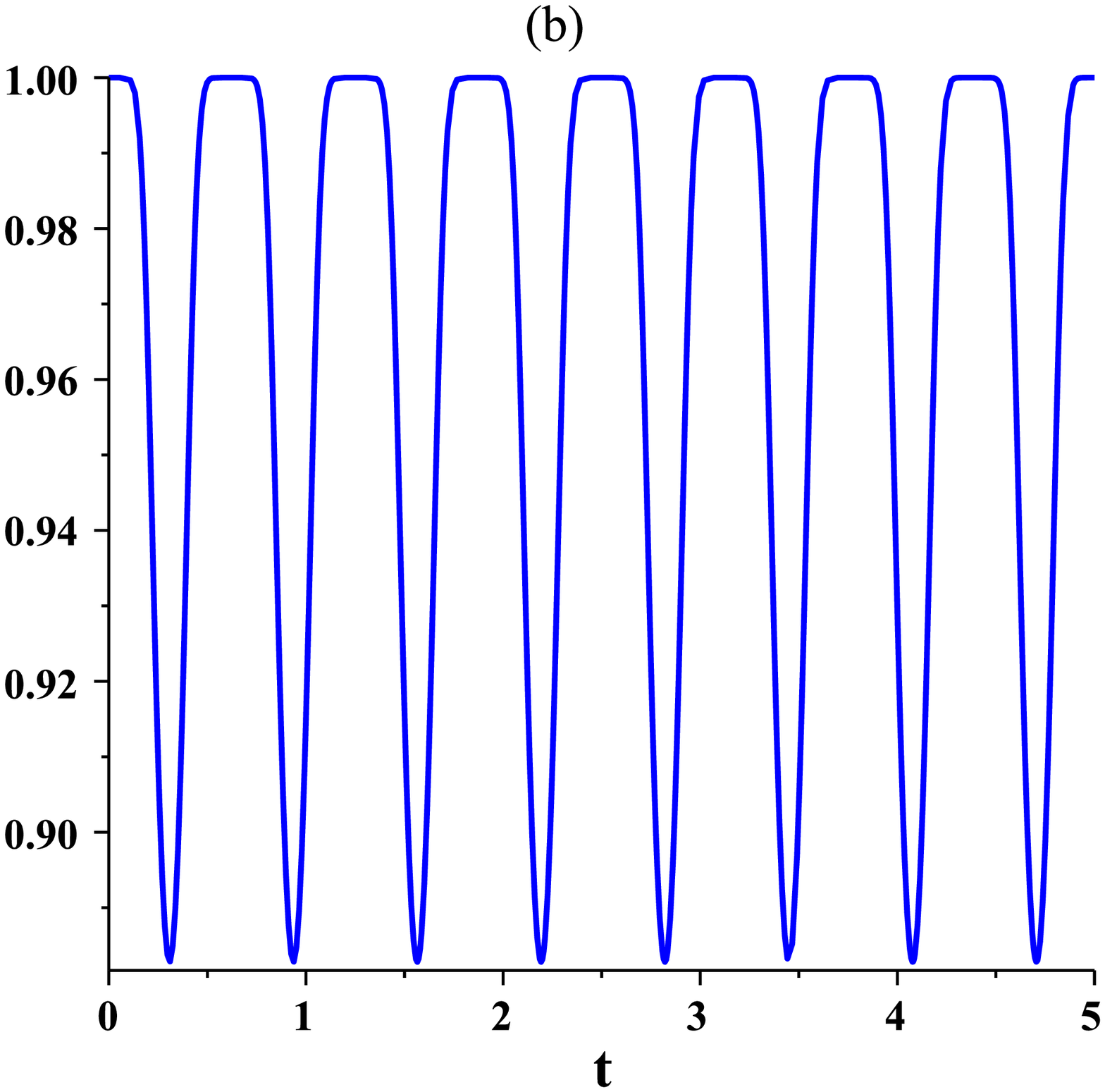}}\\
   \subfigure{\label{fig:11c}\includegraphics[width=5.5cm]{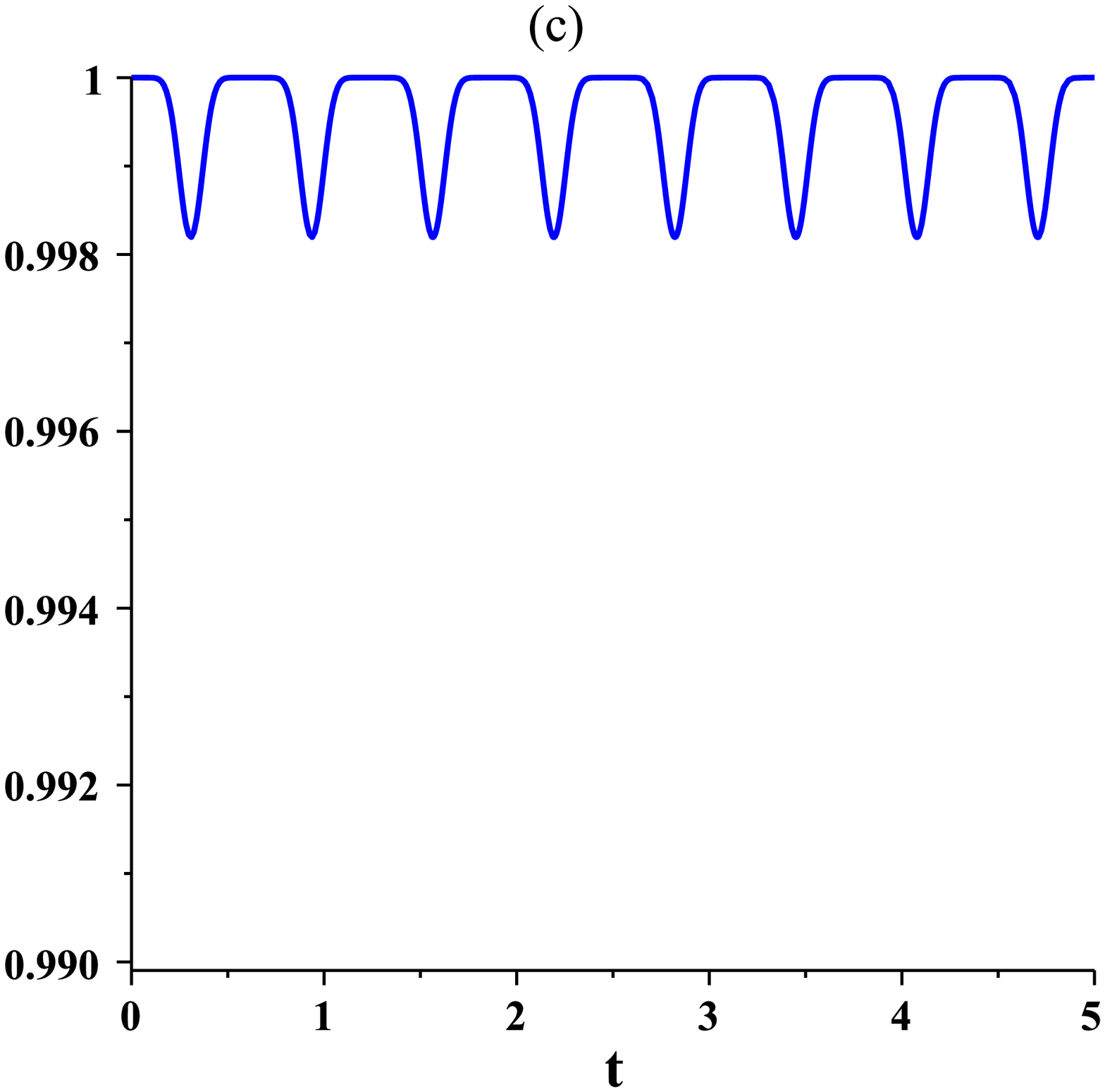}}
   \caption{{\protect\footnotesize Second integrability condition: Time evolution of concurrence of the state ${1\over \sqrt{2}}\left(|++\rangle+|--\rangle\right)$ for fixed Heisenberg interaction, $(\la_m(t))$, frequency $\beta_{m}=10$, $\kappa=0.1$, $\theta_{10}=\pi/4$ and initial phase $\phi_{m}={\pi/50}$ at different amplitudes   (a) $\mu_{m}=25 $;  (b)  $\mu_{m}=10 $;  (c)  $\mu_{m}=4 $.}}
 \label{fig:11}
 \end{minipage}
\end{figure}
Now we consider the time evolution of the concurrence for the maximally entangled state $\frac{1}{\sqrt{2}}\left(|++\rangle + |--\rangle\right)$.
We study the time evolution of the concurrence in fig.~\ref{fig:11} for fixed Heisenberg interaction, $(\la_m(t))$, frequency $\beta_{m}=10$, $\kappa=0.1$, $\theta_{10}=\pi/4$ and initial phase $\phi_{m}={\pi/50}$ at different amplitudes $\mu_{m}=25$, $10$ and $4$. As $\mu_{m}$ decreases the amplitude of the concurrence oscillation decreases as well.

\begin{figure}[htbp]
\begin{minipage}[c]{\textwidth}
 \centering
   \subfigure{\label{fig:13a}\includegraphics[width=5.5cm]{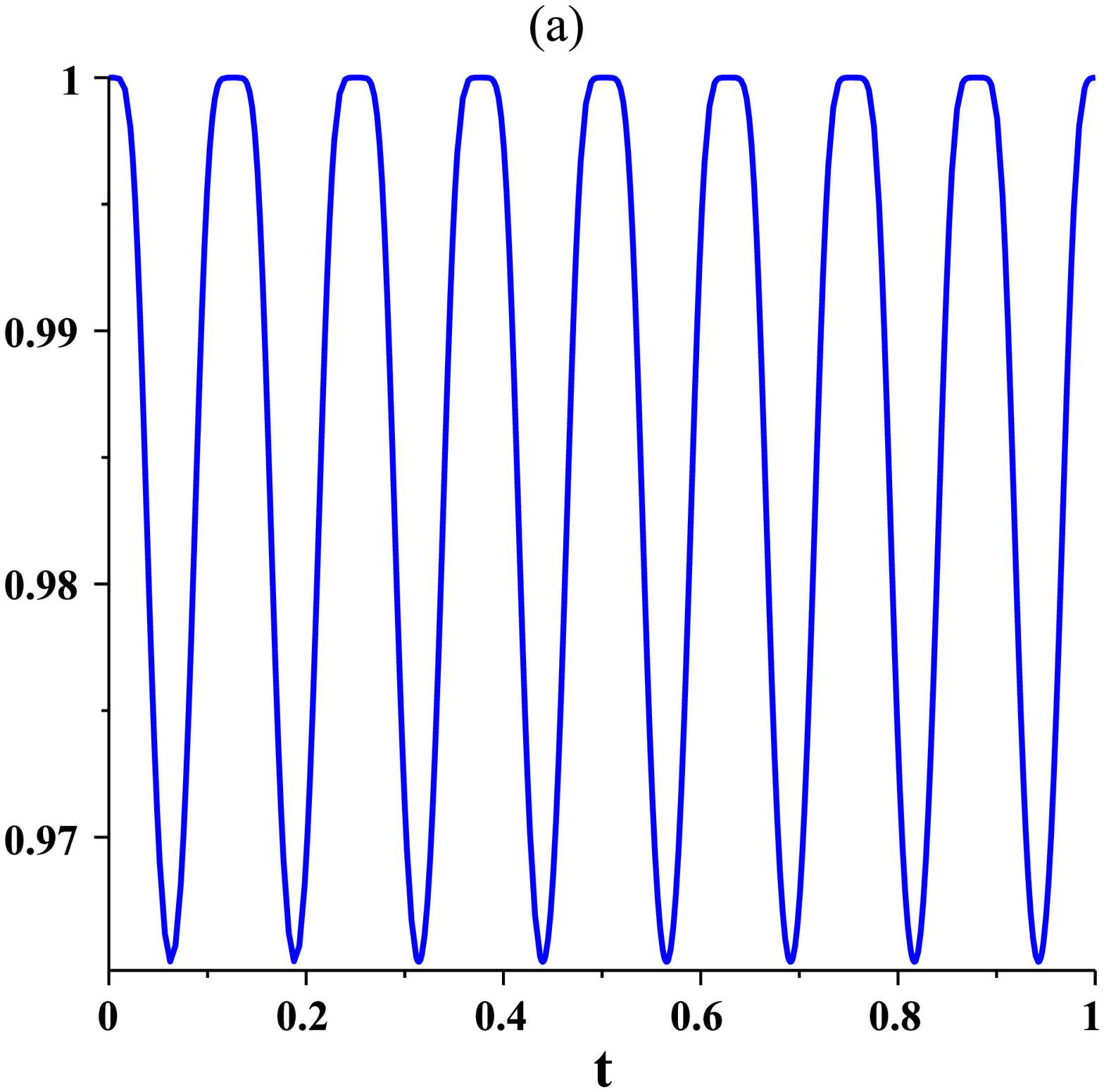}}\quad
   \subfigure{\label{fig:13b}\includegraphics[width=5.5cm]{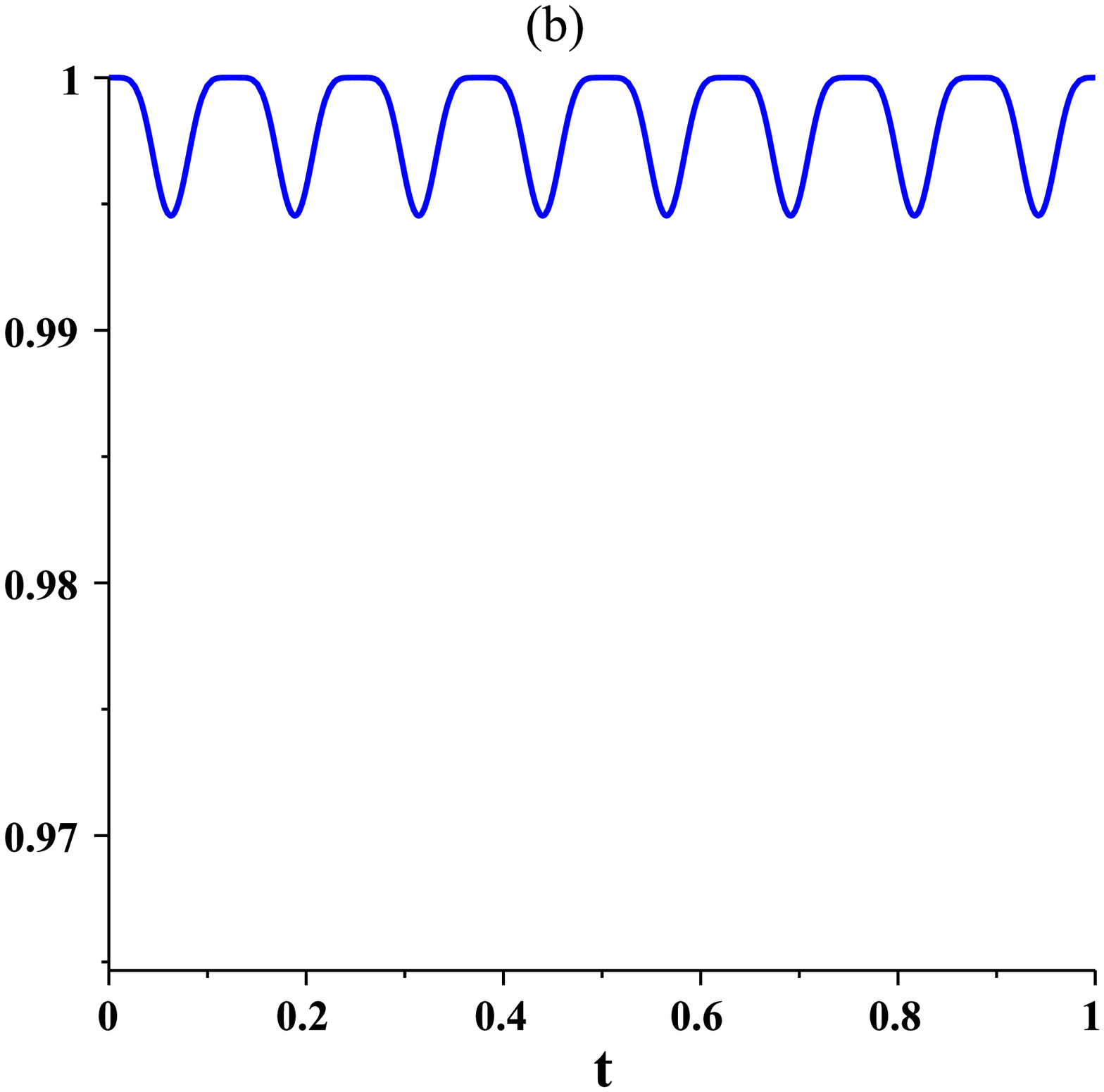}}
   \caption{{\protect\footnotesize  Time evolution of concurrence of the state $1/\sqrt{2}\left(|++\rangle+|--\rangle\right)$ for fixed Heisenberg interaction, $(\la_m(t))$,  amplitude $\mu_{m}=4$, frequency $\beta_{m}=50$ , $\theta_{10}=0$ and initial phase $\phi_{m}=0$ at  (a) $\kappa=0.1$;  (b)$\kappa=2$.}}
 \label{fig:13}
 \end{minipage}
\end{figure}

Examining the effect of different applied frequencies on the time evolution of the concurrence showed the same behavior as in the disentangled state. The frequency of the concurrence oscillation increases while its amplitude decreases as $\beta_{m}$ increases.
In fig.~\ref{fig:13} we examine the time evolution of concurrence for fixed Heisenberg interaction  amplitude $\mu_{m}=4$, frequency $\beta_{m}=50$ , $\theta_{10}=0$ and initial phase $\phi_{m}=0$ at $\kappa=0.1$ and $2$. As one can notice, increasing the magnitude of the parameter $\kappa$ leads to a smaller amplitude of the concurrence oscillation.

As to the investigation concerning the concurrence corresponding to the RWA we refer to \cite{dynamics} where a detailed investigation and study were carried out.

\section{conclusion}
In this paper we have investigated a system of two qubits coupled through a time-dependent $XYZ$ Heisenberg exchange interaction. The dynamics of the system subject to an external non-uniform time-varying magnetic field is studied. Exact solutions are provided for two different integrability conditions satisfied by the Hamiltonian. The discrete symmetries of the Hamiltonian which is responsible for dividing its Hilbert space into two independent subspaces can be utilized to map the dynamics of the system from one subspace to the other.
The time evolution of the entanglement of the system starting from different initial states, disentangled and maximally entangled, was evaluated and its different properties were investigated, which were found to be controllable using the interplay of the different parameters of the time-varying magnetic field and Heisenberg exchange interaction such as the amplitudes, frequencies and initial phases. 
\section*{Acknowledgments}
We are grateful to the Saudi NPST for support (project no. 11-MAT1492-02).


\begin{thebibliography}{99}
\bibitem{PeresBook}
A. Peres, "Quantum Theory: Concepts and Methods", (Kluwer Academic Publisher, The Netherlands, 1995).

\bibitem{Nielsen}
M. A. Nielsen and I. L. Chuang, {\it Quantum Computation and Quantum
Information} (Cambridge University Press, Cambridge, 2000).

\bibitem{Bouwmeester}
{\it The physics of Quantum information: Quantum Cryptography, Quantum Teleportation,
Quantum Computing}, Edited by D. Boumeester, A. Ekert, and A. Zeilinger
(Springer, Berlin, 2000).

\bibitem{gruska} J. Gruska, {\it Quantum Computing}
(McGraw-Hill, 1999).

\bibitem{macchiavelleo} C. Machhiavello, G.M. Palma and Z. Zeilinger,
{\it Quantum Computation and Quantum Information Theory}
(World Scientific New Jersey 2000).

\bibitem{Shor}
P. W. Shor, in Proceedings of the 35th Annual Symposium on Foundations of Computer Science,
S. Goldwasser, Ed. (IEEE Computer Society Press, Los Alamitos, CA, 1994).

\bibitem{Grover}
L. K. Grover, Phys. Rev. Lett. 79, 325 (1997).

\bibitem{Barenco}
Adriano Barenco, David Deutsch, Artur Ekert and Richard Jozsa, Phys. Rev. Lett. {\bf 74}, 4083 (1995).

\bibitem{ibm-stanford}
L. M. K. Vandersypen, Matthias Steffen, Gregory Breyta, Costantino S. Yannoni, Mark H. Sherwood, and Isaac L. Chuang ,Nature {\bf 414}, 883 (2001).

\bibitem{NMR1}
Isaac L. Chuang, Neil Gershenfeld, and Mark Kubinec, Phys. Rev. Lett. {\bf 80}, 3408 (1998).

\bibitem{NMR3}
J. A. Jones, M. Mosca, and R. H. Hansen, Nature  {\bf 393}, 344 (1998).

\bibitem{TrappedIones}
J. I. Cirac and P. Zoller, Phys. Rev. Lett. {\bf 74}, 4091 (1995); C. Monroe, D. M. Meekhof,
B. E. King, W. M. Itano, and D. J. Wineland, 75, 4714 (1995).

\bibitem{CvityQED}
Q. A. Turchette, C. J. Hood, W. Lange, H. Mabuchi, and H. J. Kimble, Phys. Rev. Lett. {\bf 75}, 4710 (1995).

\bibitem{JosephsonJunction}
D. V. Averin, Solid State Commun. 105, 659 (1998); A. Shnirman, G. Schon, and Z. Hermon, Phys. Rev. Lett. {\bf 79}, 2371 (1997).

\bibitem{trapped-ions}
J. Chiaverini, J. Britton, D. Leibfried, E. Knill, M. D. Barrett, R. B. Blakestad, W. M. Itano, J. D. Jost, C. Langer, R. Ozeri, T. Schaetz, and D. J. Wineland, Science {\bf 308}, 997 (2005).

\bibitem{supercond-junc}
D. Vion, A. Aassime, A. Cottet, P. Joyez, H. Pothier, C. Urbina, D. Esteve, and M. H. Devoret, Science {\bf 296}, 886 (2002).

\bibitem{experim-Johnson}
A.C. Johnson, J. R. Petta, J. M. Taylor, A. Yacoby, M. D. Lukin, C. M. Marcus, M. P. Hanson and A. C. Gossard, Nature  {\bf 435}, 925 (2005).

\bibitem{experim-Koppens}
F.H. L. Koppens, J. A. Folk, J. M. Elzerman, R. Hanson, L. H. Willems van Beveren, I. T. Vink, H. P. Tranitz, W. Wegscheider, L. P. Kouwenhoven, and L. M. K. Vandersypen, Science {\bf 309}, 1346 (2005).

\bibitem{experim-Petta}
J.R. Petta, A. C. Johnson, J. M. Taylor, E. A. Laird, A. Yacoby, M. D. Lukin, C. M. Marcus, M. P. Hanson, and A. C. Gossard, Science  {\bf 309}, 2180 (2005).

\bibitem{proposed-scheme}
M. Blaauboer and D.P. DiVincenzo, Phys. Rev. Lett {\bf 95}, 160402 (2005).

\bibitem{spin-qubit}
C. F. Destefani, Sergio E. Ulloa, and G. E. Marques, Phys. Rev. B {\bf 70}, 205315 (2004).

\bibitem{spin-qgate}
D. Loss and D.P. DiVincenzo, Phys Rev. A {\bf 57}, 120 (1998).

\bibitem{spin-orbit}
G. Burkard, Daniel Loss and David P. DiVincenzo, Phys Rev. B {\bf 59}, 2070 (1999).

\bibitem{nuclear-spins}
B. E. Kane, Nature (london) {\bf 393}, 133 (1998).

\bibitem{optical-lattices}
A. Sorensen et al., Nature (London) {\bf 409}, 63 (2001); W. M. Liu, W. B. Fan, W. M. Zheng, J. Q. Liang and S. T. Chui, Phys. Rev. Lett. {\bf 88}, 170408 (2002).

\bibitem{thermal-entanglement}
X. Wang, Phys. Lett. A {\bf 281} (2–3) 101 (2001).

\bibitem{spin-squeezing}
X. Wang, Phys. Rev. A {\bf 64} 012313 (2001).

\bibitem{phase-tansition}
X. Wang, Phys. Rev. A {\bf 66} 034302 (2002).

\bibitem{chain}
X. Wang, H. Fu, A.I. Solomon, J. Phys. A {\bf 34} 11307 (2001).

\bibitem{spin-chain}
M. Asoudeh, V. Karimipour, Phys. Rev. A {\bf 71} 022308 (2005).

\bibitem{disturbance}
G.-F. Zhang, S.-S. Li, Phys. Rev. A {\bf 72} 034302 (2005).

\bibitem{bendular}
Z. Huang, S. Kais, Phys. Rev. A {\bf 73} 022339 (2006).

\bibitem{mean-field}
M. Asoudeh, V. Karimipour, Phys. Rev. A {\bf 73} 062109 (2006).

\bibitem{information}
R. Rossignoli, C.T. Schmiegelow, Phys. Rev. A {\bf 75} 012320 (2007).

\bibitem{dynamics}
M.S. Abdalla, E. Lashin, G. Sadiek, J. Phys. B. {\bf 41} 015502 (2007).

\bibitem{Sadiek2010-PRA82}
G. Sadieq, B. Alkurtass and O. Aldossary, Phys. Rev. A {\bf 82} 052337 (2010).

\bibitem{Sadiek2011-PRA83}
Q. Xu, G. Sadiek and S. Kais, Phys. Rev. A {\bf 83} 062312 (2011).

\bibitem{Sadiek2012-PRA84}
B. Alkurtass, G. Sadiek and O. Aldossary, Phys. Rev. A {\bf 84} 022314 (2011).

\bibitem{multi}
Y. Sun, Y. Chen, H. Chen, Phys. Rev. A {\bf 68} 044301 (2003).

\bibitem{anisotropic}
Z.-N. Hu, S.H. Youn, K. Kang, C.S. Kim, J. Phys. A {\bf 39} 10523 (2006).

\bibitem{traveling}
A. Abliz, H.J. Gao, X.C. Xie, Y.S. Wu, W.M. Liu, Phys. Rev. A {\bf 74} 052105 (2006).

\bibitem{coupled}
G. Sadiek, Nuovo Cimento. B {\bf 125} 12 (2010).

\bibitem{solid}
G. Burkard, D. Loss, D.P. DiVincenzo, Phys. Rev. B {\bf 59} 2070 (1999).

\bibitem{Isotropic_XY}
Yang Sun, Yuguang Chen and Hong Chen, Phys Rev. A {\bf 68}, 044301 (2003).

\bibitem{woott98}
W.K. Wootters, Phys. Rev. Lett. {\bf 80}, 2245 (1998).

\bibitem{RWA}
M. Sargent, M.O. Scully, W.E. Lamb Jr., Laser Physics, (Addison-Wesley
Publishing Company, 1974).
\end{thebibliography}
\end {document}